\documentclass[11pt]{article}

\usepackage{geometry}  
\usepackage[T1]{fontenc}  
\usepackage[utf8x]{inputenc}  
\usepackage{lineno}  
\usepackage{fancyhdr}  
\usepackage{enumitem}  
\usepackage{hyperref}  
\usepackage{titling}  
\usepackage[square,numbers,sort&compress]{natbib} 
\usepackage{mathtools}  
\usepackage{titlesec}  
\usepackage{lastpage}  
\usepackage{tabularx}
\usepackage{booktabs}
\usepackage[english]{babel}  
\usepackage{amsmath}  
\usepackage{wasysym}  
\usepackage{bbm}  
\usepackage{array}  
\usepackage{xr}  
\usepackage{verbatim} 
\usepackage{float} 
\usepackage{xcolor}
\usepackage{multirow}
\usepackage{enumitem}

\usepackage{amsmath,amssymb,amsfonts}
\usepackage{algorithm}
\usepackage{algorithmic}
\usepackage{graphicx}
\usepackage{textcomp}

\newcommand{\norm}[1]{\left\lVert#1\right\rVert}

\definecolor{blue1}{RGB}{79, 113, 190}
\definecolor{orange1}{RGB}{239, 137, 51}
\definecolor{graynew}{gray}{0.88}
\definecolor{purple1}{RGB}{112, 48, 160}

\hypersetup{%
  pdfborderstyle={/S/U/W 1}
}

\setcitestyle{round,numbers}

\newcommand{\email}[1]{\href{mailto:{#1}}{{#1}}}

\newcommand{\keywords}[1]{\textbf{Keywords}: {#1}}

\newcommand{\optincludegraphics}[2][]{}

\newcommand{\optinput}[1]{}

\newtagform{brackets}{[}{]}
\usetagform{brackets}

\newcommand{\thejournal}[1]{Magnetic Resonance in Medicine}

\title{Diffusion Probabilistic Generative Models for Accelerated, in-NICU Permanent Magnet Neonatal MRI \textcolor{red}{(Submitted to Magnetic Resonance in Medicine)}}

\immediate\write18{texcount -sum=1,1,1,1,1,1,1 -merge -incbib -dir -utf8 \jobname.tex > \jobname.wcTotal }
\immediate\write18{texcount -sub=none -merge -incbib -dir -utf8 \jobname.tex | grep _main_ | grep -oE '[0-9]+' | python -c "import sys; print(sum(int(l) * w for l, w in zip(sys.stdin, [1, 1, 0, 0, 0, 0, 0])))" > \jobname.wcManuscript }
\immediate\write18{texcount -sub=none -merge -incbib -dir -utf8 \jobname.tex | grep Abstract | grep -oE '[0-9]+' | python -c "import sys; print(sum(int(l) * w for l, w in zip(sys.stdin, [1, 1, 0, 0, 0, 0, 0])))" > \jobname.wcAbstract }

\newcommand{\wcTotal}{\clearpage{\noindent\large{\bf Detailed Word Count} (not to be included for submission)}\verbatiminput{\jobname.wcTotal}}
\newcommand{\wcManuscript}{\input{\jobname.wcManuscript}}
\newcommand{\wcAbstract}{\input{\jobname.wcAbstract}}

\begin{document}

\begin{titlepage}
{\noindent\LARGE\bf \thetitle}

\bigskip

\begin{flushleft}\large
	Yamin Arefeen\textsuperscript{1,2,*},
	Brett Levac\textsuperscript{1},
    Bhairav Patel\textsuperscript{3},
    Chang Ho\textsuperscript{3},
    Jonathan I. Tamir\textsuperscript{1,3,4}
\end{flushleft}


\noindent
\begin{enumerate}[label=\textbf{\arabic*}]
\item{Chandra Family Department of Electrical and Computer Engineering\\The University of Texas at Austin\\Austin, TX, USA}
\item{Department of Imaging Physics\\MD Anderson Cancer Center\\Houston, TX, USA}
\item{Department of Diagnostic Medicine\\The University of Texas at Austin Dell Medical School\\Austin, TX, USA}
\item{Oden Institute for Computational Engineering and Sciences\\The University of Texas at Austin\\Austin, TX, USA}
\end{enumerate}

\bigskip


\textbf{*} Corresponding author:

\indent\indent
\begin{tabular}{>{\bfseries}rl}
Name		& Yamin Arefeen
\\
Department	& Chandra Family Department of Electrical Engineering and Computer Science \\
Institute	& The University of Texas At Austin														    \\
                & Austin, TX, United States
        		\\
E-mail		& \email{yaminarefeen@gmail.com}											\\
\end{tabular}\\

Word Count: $250$ (Abstract) $\sim5600$ (body)\\
\indent Figure/Table Count: 10
\end{titlepage}

\begin{abstract}
\noindent\textbf{Purpose:} Magnetic Resonance Imaging (MRI) enables non-invasive assessment of brain abnormalities during early life development. Permanent magnet scanners operating in the neonatal intensive care unit (NICU) facilitate MRI of sick infants, but have long scan times due to lower signal-to-noise ratios (SNR) and limited receive coils. This work accelerates in-NICU MRI with diffusion probabilistic generative models by developing a training pipeline accounting for these challenges.\\
\noindent\textbf{Methods:} We establish a novel training dataset of clinical, 1 Tesla neonatal MR images in collaboration with Aspect Imaging and Sha'are Zedek Medical Center.  We propose a pipeline to handle the low quantity and SNR of our real-world dataset (1) modifying existing network architectures to support varying resolutions; (2) training a single model on all data with learned class embedding vectors; (3) applying self-supervised denoising before training; and (4) reconstructing by averaging posterior samples. Retrospective under-sampling experiments, accounting for signal decay, evaluated each item of our proposed methodology. A clinical reader study with practicing pediatric neuroradiologists evaluated our proposed images reconstructed from $1.5\times$ under-sampled data.\\
\noindent\textbf{Results:} Combining all data, denoising pre-training, and averaging posterior samples yields quantitative improvements in reconstruction. The generative model decouples the learned prior from the measurement model and functions at two acceleration rates without re-training. The reader study suggests that proposed images reconstructed from $R\approx1.5$ under-sampled data are adequate for clinical use.\\
\noindent\textbf{Conclusion:} Diffusion probabilistic generative models applied with the proposed pipeline to handle challenging real-world datasets could reduce scan time of in-NICU neonatal MRI.\\
\noindent\keywords{in-NICU MRI, generative models, clinical validation, diffusion models}
\end{abstract}
\newpage

\section{Introduction}
Magnetic Resonance Imaging (MRI) of neonates enables non-invasive assessment of potential brain abnormalities during the critical phase of early life development \cite{dubois2020neonate,dudink2008neonate,ment2009neonate, woodward2006neonate, inder2013neonate}. For sick infants in the neonatal-intensive-care-unit (NICU), brain MRI helps determine the nature and extent of brain injury and potential altered brain development \cite{thiim2022neonate}. In some cases, such as infants with hydrocephalus, MRI may improve decisions on surgical intervention, compared to cranial ultrasound \cite{eldib2020neonate}. However, not all institutions acquire MRI scans for their NICU population due to limited scanner availability as very sick patients often cannot be safely transported to the 1.5 Tesla (T) or 3 T scanner rooms designed for older populations \cite{thiim2022neonate}. In response, specialized, lower-field, neonatal scanners have been designed to operate directly in the NICU \cite{thiim2022neonate,tkach2012neonate,tkach2014neonate}, allowing accommodation of the most sick and unstable NICU patients and lowering patient risk during hospital transport\cite{thiim2022neonate,bin-nun2023neonate}.

Despite the many advantages of in-NICU MRI, long scan times remain a predominant challenge and precludes access of MR imaging to many NICU patients \cite{malamateniou2013neonatee}. Sedation is not a desirable solution for helping patients tolerate lengthy scans as it increases risk to the neonatal brain and adds cost \cite{mathur2007neonate}. Most children's hospitals accomplish neonatal imaging after feeding and swaddling, promoting sleep in neonates and allowing for a successful scan. However, at least 11\% of these scans require repeat imaging due to motion, and additional techniques to shorten scan time may be beneficial \cite{janos2019neonate}, particularly in the NICU setting. In adult MRI, clinics accelerate scans \cite{zaitsev2016motionreview} with parallel imaging, which reconstructs high fidelity images from under-sampled measurements by exploiting the encoding power of the multi-coil signal receive array \cite{deshmane2012pi}. Model-based reconstruction methods, like compressed sensing, also further reduce scan time with specialized sampling patterns and hand-crafted image priors \cite{lustig2007cs,shreyas2007pcs,fessler2011modelbased}. Hospitals, vendors, and researchers also recently introduced machine learning based reconstruction algorithms to further reduce scan time in clinical settings, and these algorithms currently yield state-of-the-art accelerated MRI performance \cite{heckel2024ml}.

End-to-end machine learning methods learn a point-wise mapping between under-sampled and fully-sampled data, but are susceptible to test time shifts in the measurement operator \cite{wang2016ml,muckley2021fastmri,aggarwal2018modl,hammernik2018varnet}. Recently, researchers have applied generative methods, that learn statistical priors over a signal set of interest, for accelerated MRI acquisitions \cite{pruessmann2019bayesian,peng2020bayesian,chung2022score,song2021sde,karras2022edm,chung2023dps,luo2023diff}. As generative methods decouple the statistical prior from the measurement model, they apply in a variety of settings and do not suffer from test time shifts in the measurement operator \cite{heckel2024ml}.

The unique challenge presented by lower-field, in-NICU MRI prevents direct application of techniques designed to reduce scan times for adult and pediatric populations. Many in-NICU systems measure data with a single receive coil \cite{thiim2022neonate}, so parallel imaging cannot be applied. On the machine learning side, neural networks trained on data from adults cannot be readily applied to infants as brain anatomy varies greatly between neonates and adults, particularly in terms of incomplete myelination and gyral maturation \cite{fiagji2017neonate}. In addition, the images are inherently noisier in lower field MRI, and the permanent magnet in-NICU scanners suffer from field inhomogeneity and residual magnetization artifacts typically not seen with standard super conducting magnets used in adult MRI \cite{arnold2022neonate}. Finally, fewer public repositories of neonatal MRI data are available, making training machine learning reconstruction models for accelerated infant imaging challenging from both a data quality and quantity perspective.

To the best of our knowledge, this work presents the first diffusion-probabilistic generative model trained on real-world, lower-field, in-NICU neonatal MRI data from a diverse range of image contrasts and anatomical orientations to solve inverse problems that accelerate acquisitions and improve motion robustness. We employ generative models for their robustness to test-time shifts in the measurement operator, enabling application when in-NICU neonatal MRI requires frequent adjustment of the imaging protocol for different patients or with retrospective motion correction when each test sample experiences variation in the measurement model. Our contributions are the following:
\begin{itemize}[left=0pt,nosep]
\item We establish a novel training dataset of real-world in-NICU, clinical MRI images in collaboration with Aspect Imaging and Sha'are Zedek Medical Center acquired with Aspect Imaging's 1 Tesla Embrace system, equipped with a permanent magnet and single channel solenoid RF coil. The dataset consists of axial, sagittal, and coronal T$_2$ fast spin echo (FSE) and axial T$_1$ spin echo (SE) scans.
\item We propose a training pipeline that combines a number of machine learning methods to handle the challenges associated with our real-world dataset, including low quantity and low signal-to-noise-ratio (SNR): (\textbf{1}) we modify popular existing diffusion network architectures to support inputs with varying matrix sizes, a common variability in neonatal MRI, to expand the set of potential training images; (\textbf{2}) rather than stretching our dataset thin by training a separate model for each contrast and slice orientation, we train a single model on all data using a learned class embedding vector; (\textbf{3}) we apply self-supervised denoising as a pre-training step to boost the SNR of our dataset before training; (\textbf{4}) we reconstruct an image by approximating the conditional expectation through averaging multiple samples from the posterior distribution.
\item We present the first clinical reader study where radiologists evaluate accelerated in-NICU neonatal MR acquisitions reconstructed with generative models. Our under-sampling correctly accounts for FSE signal decay, and therefore represents a realistic retrospective acceleration experiment.
\end{itemize}

First, we quantify the impact of each item in our proposed methodology with experiments that evaluate the effect of training a single model on all data with class embeddings and experiments that investigate the effects of denoising and compare to baseline methods. In addition, we verify the benefits of averaging multiple posterior samples for reconstruction in our NICU setting \cite{luo2023diff}. This is followed by an ablation study that comprehensively compares models trained with and without class embedding and denoising and evaluates performance across varying measurement operators. Finally, we perform a reader study where practicing neuroradiologists compare conventional images reconstructed from fully-sampled data to images reconstructed with our proposed approach from $1.5\times$ under-sampled data to see if our method maintains image quality and without hallucinating or removing structure.

A preliminary version of this work appears in ISMRM 2025 \cite{yamin2025ismrm}.

\section{Theory}
We model single-channel neonatal MRI data with $\mathbf{y}= \mathbf{A} \mathbf{x} + \mathbf{\eta}$, where $y\in\mathbb{C}^m$ are the acquired measurements, $\mathbf{A}\in\mathbb{C}^{m\times n}$ is the under-sampled Fourier operator, $\mathbf{x}\in\mathbb{C}^n$ represents the vectorized clean image, and $\mathbf{\eta}$ is Gaussian measurement noise with standard deviation $\sigma_d$. When reconstructing from under-sampled data for reduced scan time ($m < n$), prior knowledge must be incorporated to estimate a high fidelity image. Generative models \cite{bond2022DeepGM} solve ill-posed inverse problems by learning a statistical prior, $p(x)$, to guide the reconstruction towards solutions that both match the data and are statistically likely. These methods assume that the image to reconstruct is a sample from the probability distribution described by the dataset used to train the generative model.

Specifically, score-based diffusion models indirectly learn $p(\mathbf{x})$ by training a neural network, $D_{\theta}$, to approximate the score $\nabla_{\mathbf{x_t}} \log p_t(\mathbf{x})$, where $p_t(\mathbf{x})$ is the original data distribution perturbed by Gaussian noise with standard deviation $\sigma_t$ \cite{song2021sde,karras2022edm,ho2020diff,yang2023diff}. Starting from a sample from the Gaussian distribution, solving a differential equation by running an iterative denoising procedure using the learned score-function yields a sample from the prior distribution $p(x)$ \cite{song2021sde}. Score functions can also be used to approximately sample from the posterior distribution, $p(x|y)$, in the context of ill-posed inverse problems \cite{jalal2021robust,chung2022score, luo2023diff}.

Following the prior sampling formulation in \cite{karras2022edm}, we sample from the posterior distribution in accelerated neonatal MRI by solving the ordinary differential equation (ODE):
\begin{equation}
    d\mathbf{x} = \left[\frac{\dot{s}(t)}{s(t)}\mathbf{x} - s(t)^2 \dot{\sigma}(t)\sigma(t)\nabla_{\mathbf{x}} \log p\left( \frac{\mathbf{x}}{s(t)} | \mathbf{y}; \sigma(t) \right)\right] dt.
\end{equation} with $s(t)=1$ and $\sigma(t)=t$. Using Baye's rule, we separate the posterior score into a log-likelihood and prior score,
\begin{equation}
    d\mathbf{x} = \left[-t \left(\nabla_\mathbf{x} \log p\left( \mathbf{y}|\mathbf{x} ; t \right) + \nabla_\mathbf{x} \log p\left( \mathbf{x} ; t \right)  \right)\right] dt.
\end{equation}Substituting the diffusion model that estimates the prior score function and the analytical expression for the likelihood score \cite{jalal2021robust,chung2023dps} yields the following ODE that we solve to obtain posterior samples,
\begin{equation}
    d\mathbf{x} = \left[-t \left(\nabla_\mathbf{x} \norm{\mathbf{A}\mathbf{\tilde{x}}(\mathbf{x}) - \mathbf{y}}_2^2 + D_\theta(\mathbf{x},t\right) \right] dt.
    \label{eq:ode}
\end{equation}The analytical expression for the likelihood score is only known at time point $t=0$ so we use the approximation $\mathbf{\tilde{x}}(\mathbf{x}) = \mathbb{E}[\mathbf{x_0}|\mathbf{x}]$ \cite{chung2023dps}. To produce a single reconstructed image, one can average multiple posterior samples obtained by solving the ODE in Equation [\ref{eq:ode}] starting from multiple Gaussian initializations \cite{luo2023diff}. Note how this formulation decouples the statistical prior from the likelihood, so for neonatal MRI, a single prior can be re-used to solve inverse problems with different sampling patterns, receive coils, timings, and measurement models.

\section{Methods}
\subsection{Neonatal Dataset}
In collaboration with Aspect Imaging and Sha'are Zedek Medical Center, under Institutional Review Board approval and informed consent/assent, we acquired a dataset with the in-NICU 1T Aspect Embrace System, equipped with a permanent magnet and single channel solenoid RF coil, on 128 clinical neonatal patients. For each subject the following pulse sequences were acquired: 2D T$_2$-weighted axial, sagittal, and coronal fast-spin-echo (FSE) and 2D T$_1$-weighted axial spin-echo (SE). Due to severe patient motion, clinical limitations, and differences in anatomy, no single standardized protocol was used to scan all subjects. We split the dataset into 108 subjects for training and validation and 20 subjects for testing, resulting in 8,659 FSE and 3,224 SE training slices. 

Both FSE and SE protocols selected from discrete readout and phase encode field-of-view values of [$110,115,120,130,140$] mm. The FSE protocols had an average readout resolution of $0.8\pm 0.1$ mm and phase-encode resolution of $0.8\pm 0.1$ mm. FSE scans also employed echo train lengths from the discrete set of values \{$12,14,18,20,22$\} with an average echo spacing of $11.3\pm 1.0$ ms, average echo time of $147.1\pm 12.0$ ms and repetition time of $7125\pm 1061$ ms. The SE scans had average readout and phase-encode resolutions of $0.9\pm 0.08$ mm and $0.9\pm 0.09$ mm respectively. SE also used single echo measurements per repetition, with an average echo time of $10.4\pm 0.6$ ms and repetition time of $629\pm 45$ ms. Both FSE and SE measured between [$16-30$] slices with a thickness of $3.1\pm 0.3$ mm and $3.1\pm 0.4$ mm and gap of $1.0\pm 0.4$ mm and $1.0\pm 0.6$ mm respectively. FSE and SE had average scan times of $69.6\pm 8.4$ s and $94.8\pm10.9$ s respectively. Standard main field \cite{aspectb0} and transmit field inhomogeneity correction available on the Embrace System were not applied to our training dataset.

\subsection{Proposed Methodology for Generative Diffusion Model Training and Image Reconstruction}

Naive application of standard generative model training methods do not work well in our in-NICU neonatal MRI setting due to the low quantity of and corruptions in the data, therefore we developed a training and reconstruction pipeline to address these problems, summarized in Figure 1. 

\begin{figure*}[t!]
    \centering
    \centerline{\includegraphics[width=\linewidth]{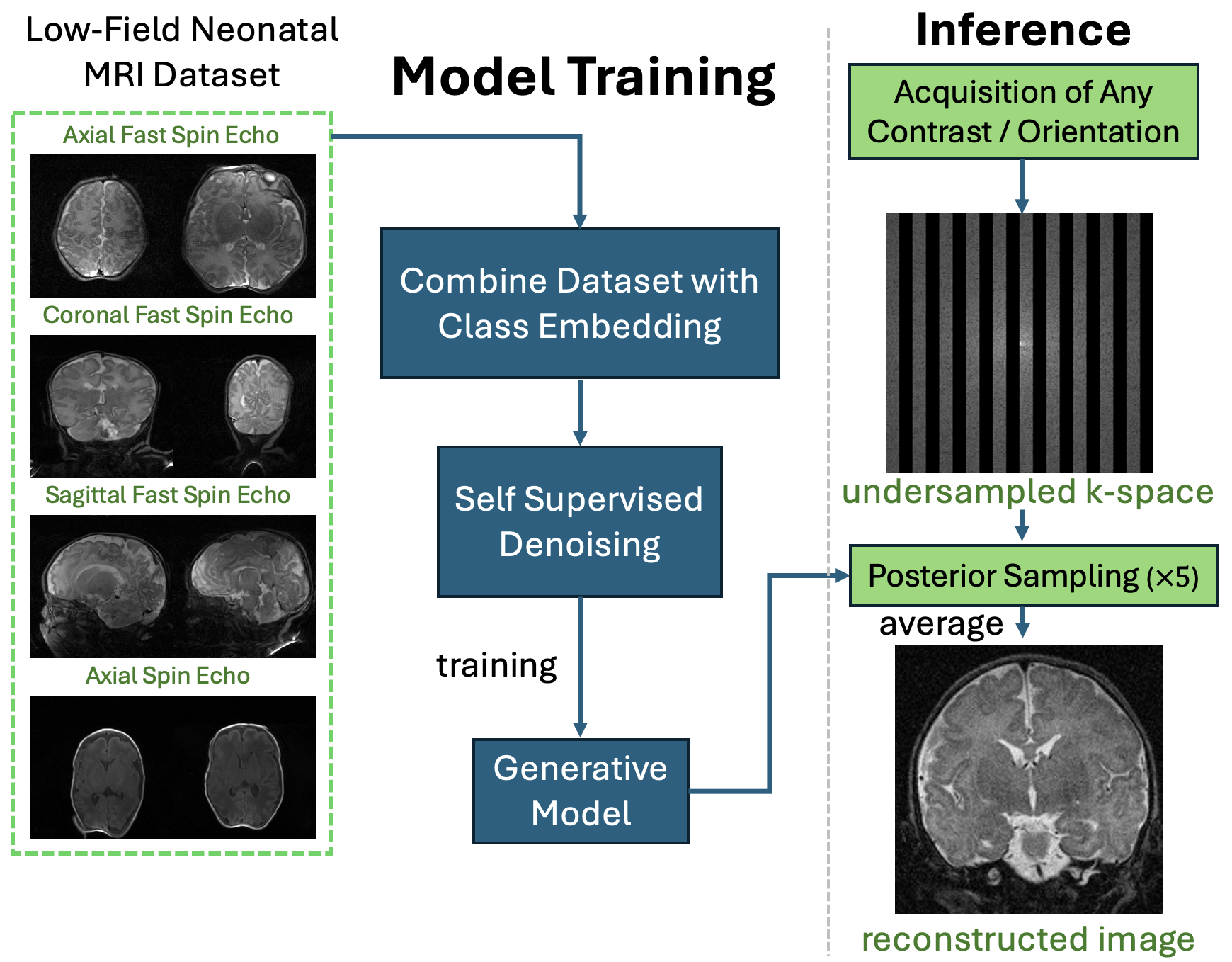}}
    \caption{The proposed training and inference pipeline. Images from different orientations and contrasts are combined into a single dataset using class embeddings. A denoiser, trained in a self-supervised fashion, denoises the dataset before training. Diffusion posterior sampling reconstructs an image from under-sampled k-space by averaging 5 posterior samples. The proposed methods enable training of diffusion models for accelerated reconstruction on noisy and limited in-NICU neonatal MRI data.}
\end{figure*}

\subsubsection{Existing Diffusion Model Architecture and Proposed Modifications}
We based our generative model on an implementation of a U-Net style architecture from Ref.\cite{song2021sde} provided by Ref.\cite{karras2022edm} with approximately 65 million parameters. For training, we followed the "EDM" loss formulation, data augmentation, learning rate schedule, optimizer choice, and noise schedule with a batch size of 180 parallelized across four A100 GPUs \cite{karras2022edm}. The existing model only takes inputs with a single, pre-prescribed matrix size, but in-NICU neonatal MRI consists of data acquired with a range of matrix sizes, as evidenced by the variability in resolution and field-of-view in our dataset. To account for this, we modify the down-sampling and up-sampling portions of the U-net. We set pre-determined resolutions at each U-Net level, based on the dimensions of the training data, and then perform the up-sampling or down-sampling with bi-linear interpolation to those resolutions. In this way, the model can take inputs with arbitrary matrix sizes. We re-size all of training data to matrix size $200 \times 200$ for model training, and then we can apply the model to heterogeneously sized data during inference.

\subsubsection{Training on All Data with Class Embeddings}
While our total dataset consists of 11,883 training slices, splitting it along contrast and orientation yields 3,599 axial FSE, 2,687 coronal FSE, 2,373 sagittal FSE slices, and 3,244 axial SE slices. Assuming a different statistical prior, p(x), for each contrast and orientation and training separate models for each spreads the limited data too then. On the other hand, training on all data together improves model robustness while maintaining performance \cite{lin2023robustness}, but this assumes that all images come from the same prior distribution, despite the marked differences in image content between the contrasts and orientations. To address both these limitations, we train on all data simultaneously with learned embedding vectors that condition the model based on the current image contrast and orientation being sampled \cite{karras2022edm}.

We define four classes for our dataset, with each class assigned a one-hot encoded vector: (i) FSE Axial: $[1, 0, 0, 0]$ (ii) FSE Coronal: $[0, 1, 0, 0]$ (iii) FSE Sagittal: $[0, 0, 1, 0]$ (iv) SE Axial: $[0, 0, 0, 1]$. When taking an input from a particular class, the network takes the corresponding one-hot encoded vector through a relatively small fully-connected neural network to produce an embedding vector with $512$ dimensions. The neural network that produces the embedding vector is trained simultaneously with the generative model. The embedding vector then serves as an additional input to each of the U-Net blocks in the larger model. Since the image class is known apriori for an accelerated MRI acquisition, the embedding vector can be incorporated into posterior sampling, enabling the model to take advantage of class information during reconstruction while also being trained on all available data.

\subsubsection{Self-supervised Denoising before Training}
When training models on data from higher field scanners with super-conducting magnets, the measurement noise, $\eta$, is typically assumed to be negligible. However, a combination of the lower-field and permanent magnet makes noise in our in-NICU data non-negligible. Previous work has shown that denoising training data improves performance of learning based reconstruction methods \cite{asad2025denoising}, so we train another U-net to denoise our dataset before generative model training. The denoising U-net consists of the same architecture as the diffusion model, with 65 million parameters. Since we only have access to our inherently noisy dataset, we train the denoiser in a self-supervised fashion using Noisier2Noise with slight modifications \cite{moran2020n2n}. Each time the model sees a training sample for loss calculation, we use a small $20 \times 20$ patch on the background of the image to estimate its noise standard deviation. Then Gaussian noise, with $1.5 \times$ the estimated standard deviation, is added to the training sample. The model is applied to this noisier sample, and a L$_2$ loss is computed between the original and noisier sample. After the model has been trained, we apply it to the original dataset for denoising before generative model training. We used the Adam optimizer with the "EDM" learning rate scheduler \cite{karras2022edm}.

\subsubsection{Diffusion Posterior Sampling Details}
To perform posterior sampling with our diffusion models, we combine the prior sampling procedure from \cite{karras2022edm} with Diffusion Posterior Sampling \cite{chung2023dps}. Algorithm \ref{alg:dps} details our image reconstruction procedure given under-sampled data, $y$, the associated one-hot encoding vector that labels the class of the data, $C$, and the diffusion model $D_{\theta}$. Five averaged posterior samples, to estimate the conditional expectation, gives our reconstructed image. We fixed reconstruction hyperparameters across all experiments.

\begin{algorithm}
\caption{Image Reconstruction with Diffusion Posterior Sampling}\label{alg:dps}
\begin{algorithmic}
    \REQUIRE $D_{\theta}, y, C$
    \STATE $\sigma_{\min} \gets 0.002, \sigma_{\max} \gets 5.0, \rho \gets 7.0, N_t \gets 450, N_s \gets 5$
    \STATE Initialize $\mathbf{x} \gets 0$
    \FOR{$s \gets 1$ to $N_s$}
        \STATE Initialize $\mathbf{x_0} \gets N(0,\sigma_{\max})$
        \FOR{$i \gets 1$ to $N_t$}
            \STATE $t_i \gets [\sigma_{\max}^{1/\rho} + \frac{i (\sigma_{\min}^{1/\rho}-\sigma_{\max}^{1/\rho})}{N_t}]^{\rho}$
            \STATE $\mathbf{\hat{x}_{i-1}} = D_{\theta}(\mathbf{x_{i-1}},C)$
            \STATE $\mathbf{d_P} = \frac{\mathbf{x_{i-1}}-\mathbf{\hat{x}_{i-1}}}{t_i}(t_{i+1}-t_i)$
            \STATE $r = ||\mathbf{A}\mathbf{\hat{x}_{i-1}}-\mathbf{y}||_2^2$
            \STATE $\mathbf{d_L} = -\frac{\nabla_{\mathbf{x_{i-1}}}(r)}{\sqrt{r}}$
            \STATE $\mathbf{x_i} = \mathbf{x_{i-1}}+\mathbf{d_P}+\mathbf{d_L}$
        \ENDFOR
        \STATE $\mathbf{x} = \mathbf{x} + \mathbf{x_{N_t}}$
    \ENDFOR
    \RETURN $\mathbf{x} / N_s$
\end{algorithmic}
\end{algorithm}

\subsection{Quantitative Under-sampling Experiments}
All quantitative experiments retrospectively under-sampled test examples by removing data associated with a specific echo train as a group to capture realistic signal decay \cite{rajput2024retro,FSE_comp_moco,shimron2022ImplicitDC}. To quantitatively evaluate the various reconstruction methods, normalized-root-mean-squared-error (NRMSE) was computed with respect to the fully-sampled image. 

\subsubsection{Training with All Data using Class Embeddings}
We compared the reconstruction performance of a generative diffusion model trained on all data with and without class embeddings to models trained separately for each contrast and orientation class. Test data were retrospectively under-sampled along the phase-encode direction by a factor of $R=2$. The test dataset consisted of 220 axial FSE, 110 coronal FSE, 110 sagittal FSE, and 210 axial SE slices. No model in this experiment employed denoising before training.

 In addition we compared reconstructions with the aforementioned generative models to MoDL \cite{aggarwal2018modl}, a state-of-the-art end-to-end method for accelerated MRI reconstruction. MoDL models were either trained on each contrast and orientation class separately or trained on all data simultaneously. Each MoDL model had 65 million trainable parameters, to match the generative models, with 6 unrolls consisting of data consistency followed by UNet-based regularization. 

\subsubsection{Effect of Denoising and Comparison to Non-learned Baseline}
Next, we compared a non-learned L1-wavelet based reconstruction, implemented with BART \cite{bart}, to our generative diffusion model trained with and without denoising pre-training. Both generative models were trained on all data with class embeddings. Test data were retrospectively under-sampled by an average factor of $R=1.5$. The test data for this experiment consisted of 110 axial FSE, 80 coronal FSE, 60 sagittal FSE, and 110 axial SE slices.

\subsubsection{Performance from Averaged Posterior Samples}
Reconstruction performance was compared when averaging \{$1,2,3,4,5$\} posterior samples from a generative model trained on all data with class embedding and denoising. The aforementioned test dataset, with an average acceleration rate of $R=1.5$, and 110 axial FSE, 80 coronal FSE, 60 sagittal FSE, and 110 axial SE slices was used.

\subsubsection{Denoising and Class Embedding Ablation on Two Measurement Models}
Finally, we analyzed the combined effects of denoising and using class embeddings by training four models on all data: (i) without embeddings, without denoising (ii) without embeddings, with denoising (iii) with embeddings, without denoising (iv) with embeddings, with denoising. Test data were retrospectively under-sampled by an average factor of $R=\{1.5,2.0\}$, and the generative models were applied to both under-sampling rates without any modifications. Test data consisted of 240 axial FSE, 110 coronal FSE, 120 sagittal FSE, and 110 axial SE slices.

\subsection{Reader Study}
To analyze whether the proposed method maintains quality and anatomical delineation without hallucinating or removing structure when reconstructing images from under-sampled data, we performed a reader study with two pediatric neuroradiologists (C.H. and B.P. with $18$ and $20$ years of experience respectively). 

\subsubsection{Evaluation 1: Blind Comparison of Volumes}
Radiologists evaluated 36 volumes, where 18 volumes were reconstructed conventionally from fully-sampled data and the other 18 volumes were the same data reconstructed with our proposed approach after retrospective under-sampling by an average $R=1.5$. The radiologists viewed the volumes in a random order and were blinded to the method used to reconstruct the images. The original 18 volumes consisted of 6 FSE axial, 5 FSE coronal, 2 FSE sagittal, and 5 SE axial scans.

The reader study first asked the radiologists to rate general image quality criteria for each volume: \textit{SNR}, \textit{Contrast-to-Noise-Ratio (CNR)}, \textit{artifact level}, and \textit{overall image quality}, on the scale: [$1 \gets$ poor | $2 \gets$ fair | $3 \gets$ good].

The readers also rated each volume for structure delineation of the \textit{white matter}, \textit{cortex}, \textit{cerebellum}, \textit{brain stem}, and \textit{grey matter} on the scale: [$1 \gets$ non-diagnostic | $2 \gets$ limited | $3 \gets$ diagnostic].

Paired, nonparametric Wilcoxon signed-rank tests evaluated if there was a statistically significant difference in ratings between fully-sampled and proposed for each criteria.

\subsubsection{Evaluation 2: Direct Paired Comparison of Slices}

In the second phase of the reader study, radiologists were shown 36 pairs of images where the first image in the pair was the conventional image reconstructed from fully-sampled data and the second image was reconstructed using the proposed approach from the same data with $R=1.5$ under-sampling. The 36 pairs consisted of 12 FSE axial, 10 FSE coronal, 4 FSE sagittal, and 10 SE axial slices.

For image quality, the radiologists were asked to evaluate \textit{SNR}, \textit{CNR}, and \textit{Overall Image Quality} on the following scale: [$-2 \gets$ conventional heavily preferred | $-1 \gets$ conventional slightly preferred | $0 \gets$ same | $1 \gets$ proposed slightly preferred | $2 \gets$ proposed heavily preferred]. An unpaired, nonparametric Wilcoxon signed-rank test evaluated whether the ratings varied significantly from $0 \gets$ same.

The radiologists were also asked: \textit{Is there a structure or features that exists in the conventional image but does not exist on the proposed image?}. The question was answered using the following scale: [$1 \gets$ proposed image misses many structures | $2 \gets$ proposed images slightly misses some structure | $3 \gets$ proposed image misses no structure]

Finally, the study asked: \textit{Is there structure or features that do not exist in the conventional image but exist in the proposed image?} The question was answered using the scale: [$1 \gets$ proposed has many additional structures | $2 \gets$ proposed image has slightly additional structures | $3 \gets$ proposed has no additional structure].

\section{Results}
\subsection{Denoising and Example Prior Samples}
Figure 2 (A) shows our denoising model, trained in a self-supervised fashion, applied to two example training slices. Qualitative noise reduction is observed, illustrated by the panels with increased brightness. Figure 2 (B, C, D, E) shows prior samples generated with our proposed model trained on all data with class embeddings and denoising pre-training. The model generates either FSE axial, FSE coronal, FSE sagittal, or SE axial samples based on the input class embedding specified at inference. In this way, we use all data to train the model, and the model can adapt to the specific class of data to reconstruct with posterior sampling.

\begin{figure*}[t!]
    \centering
    \centerline{\includegraphics[width=\linewidth]{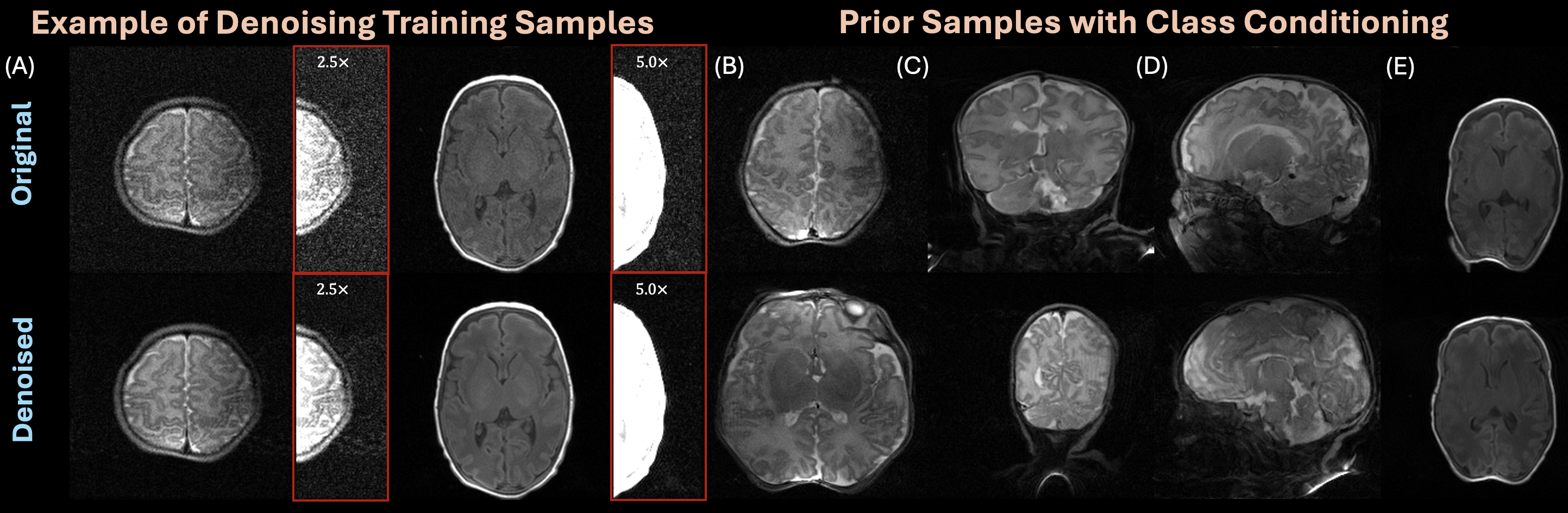}}
    \caption{(a) The top row shows two training samples from our dataset and the bottom row shows the corresponding training samples after applying our denoiser trained in a self-supervised fashion. (b,c,d,e) prior samples generated by our trained model when conditioned on class embeddings of fse axial, sagittal, coronal, and se axial. Our model uses all available training data to learn a statistical prior over neonatal MR images.}
\end{figure*}

\subsection{Quantitative Under-sampling Experimental Results}
The violin plot in Figure 3 compares reconstructions of the different image classes with models trained separately on each image class, a model trained on all data without embedding, and a model trained on all data with embedding. Both models trained on all data achieve lower NRMSE across the dataset in comparison to the models trained separately. In addition, incorporating class embedding improves performance when training on all data. Figure 4 shows example reconstructions on all four image classes from this experiment. Similar to the overall quantitative setting, the model using all data with class embeddings achieves the lowest quantitative and qualitative error.

Supporting Figure S1 compares reconstructions from the generative models with the end-to-end MoDL models trained on each image class separately or trained on all data simultaneously with violin plots, and Supporting Figure S2 shows example reconstructions on all four image classes.

\begin{figure*}[t!]
    \centering
    \centerline{\includegraphics[width=\linewidth]{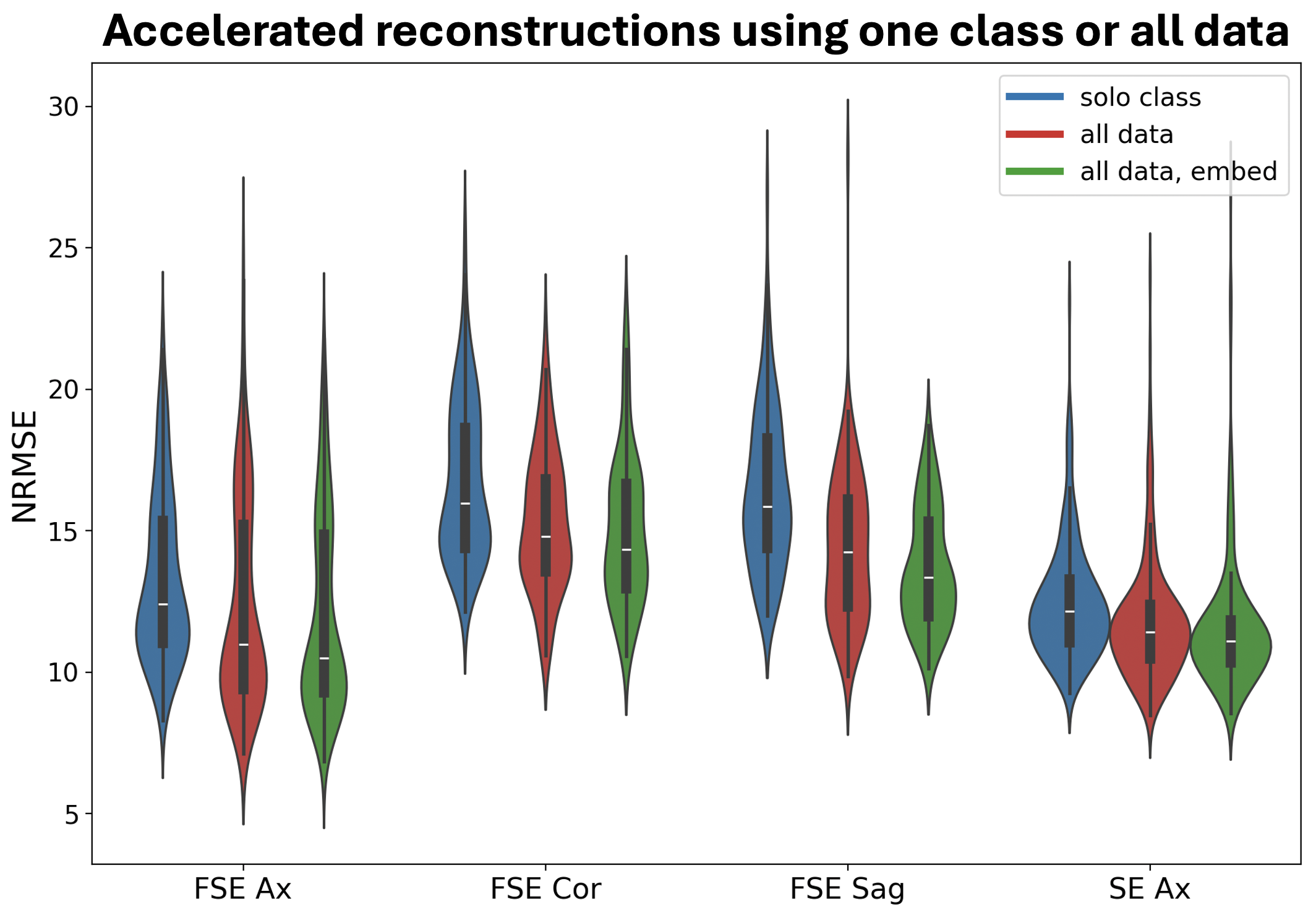}}
    \caption{Violin plots comparing NRMSE of posterior sampling reconstructions on $R=2$ under-sampled data using diffusion models trained on just a single image class, trained on all data, and trained on all data with class embeddings. Results on all contrasts and orientations suggest that a diffusion model trained by combining all data with class embeddings yields the best quantitative performance.}
\end{figure*}

\begin{figure*}[t!]
    \centering
    \centerline{\includegraphics[width=\linewidth]{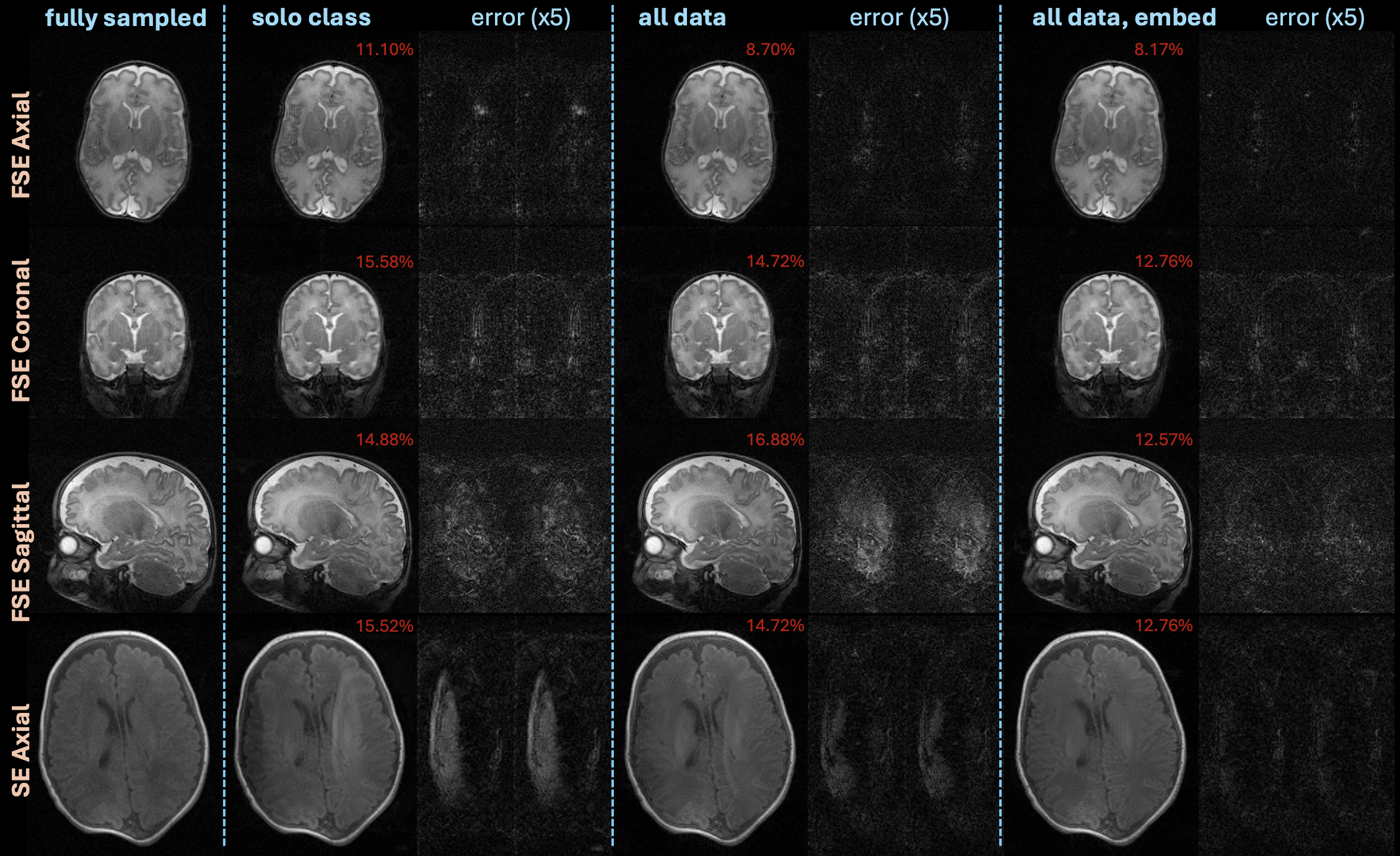}}
    \caption{Example images from the experiment comparing posterior sampling reconstructions on $R = 2$ under-sampled data using diffusion models trained on just a single image class, trained on all data, and trained on all data with class embeddings. This is a specific illustration of the quantitative conclusion that using all data combined with class embeddings to train the diffusion model yields best performance.}
\end{figure*}

The violin plot in Figure 5 compares NRMSEs of reconstructions using non-learned L$_1$-wavelet to generative models trained with class embeddings on all data with and without denoising. The non-learned method performs worse than both learned approaches, particularly  on the FSE data. The difference in quantitative performance between models trained with and without denoising is small. Supporting Figure S3 shows example reconstructions from this experiment.

\begin{figure*}[t!]
    \centering
    \centerline{\includegraphics[width=\linewidth]{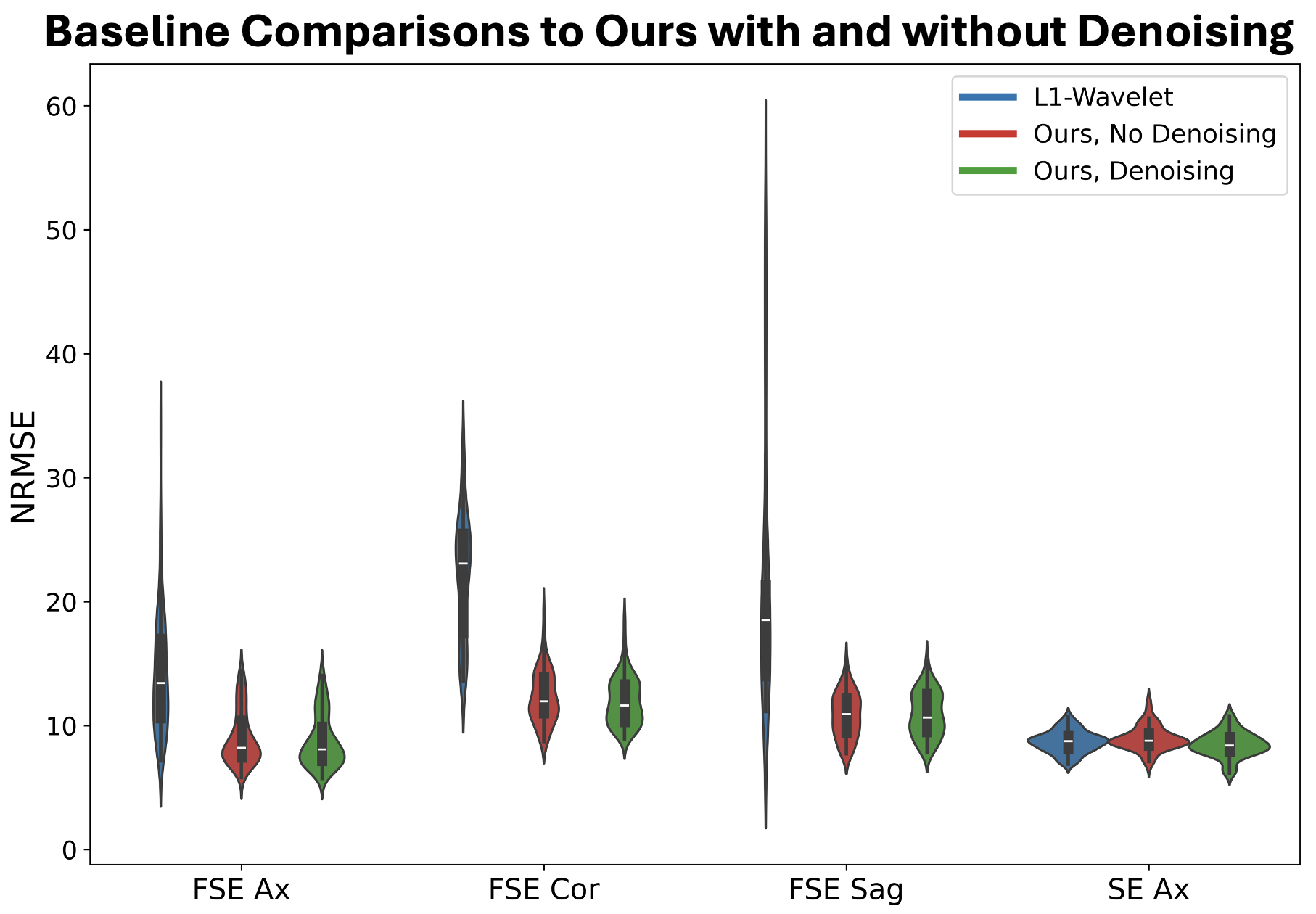}}
    \caption{Violin plots comparing NRMSE of reconstructions on $R=1.5$ under-sampled data using baseline L$_1$ and diffusion models trained on all data with and without denoising. While using a learned prior provides benefit, little quantitative difference exists between models trained with and without denoising pre-training.}
\end{figure*}

Figure 6 illustrates the effect of averaging posterior samples to produce a reconstructed image. Averaging more posterior samples reduces the NRMSE across the test set for all four image classes, as illustrated by the plot in Figure 6 (A). Figure 6 (B) shows specific FSE sagittal and SE axial examples with reconstructions averaging varying numbers of posterior samples. Averaging more posterior samples improves qualitative and quantitative performance. 

\begin{figure*}[t!]
    \centering
    \centerline{\includegraphics[width=\linewidth]{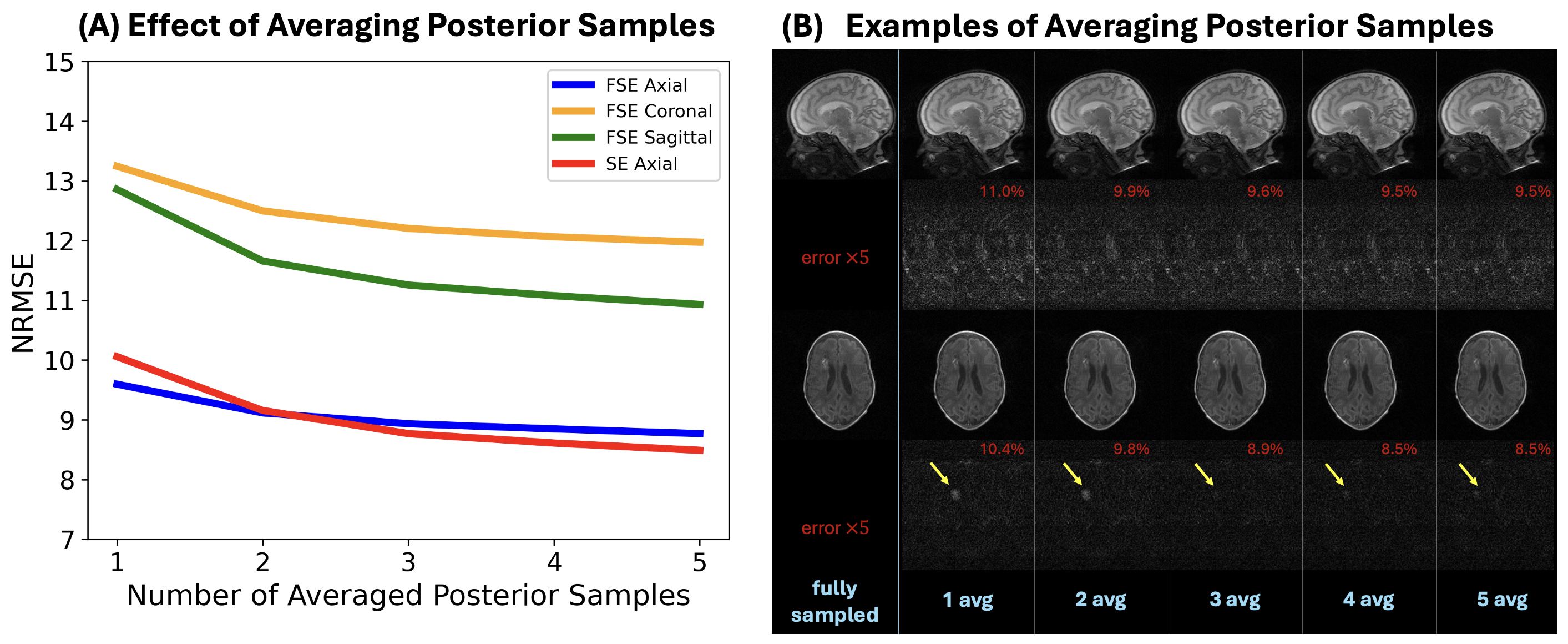}}
    \caption{(A) Mean NRMSE of a generative model trained with class embeddings on all data with pre-training denoising when averaging \{$1,2,3,4,5$\} posterior samples. Averaging more posterior sample leads to improved quantitative performance. (B) Two specific examples from the experiment where averaging more posterior samples improves quantitative and qualitative performance.}
\end{figure*}

Figure 7 shows violin plots comparing models trained on all data with and without class embeddings and denoising pre-training. In addition, the plots present results at average acceleration rates of 1.5 and 2 with no modifications to the models or posterior sampling hyperparameters. While class embedding improves quantitative performance, denoising does not yield clear improvements. Examples from the test set at $R = 2$ are shown in Figure 8. Applying embedding mostly improves quantitative and qualitative performance, but a similar trend is not observed with denoising. However, in the SE results, and its associated zoomed in row, the no embed, no denoise model achieves lower NRMSE than the no embed, denoise model, but suffers from increased hallucination artifacts, particularly in the ventricles, highlighted by the zoomed in panels.

\begin{figure*}[t!]
    \centering
    \centerline{\includegraphics[width=\linewidth]{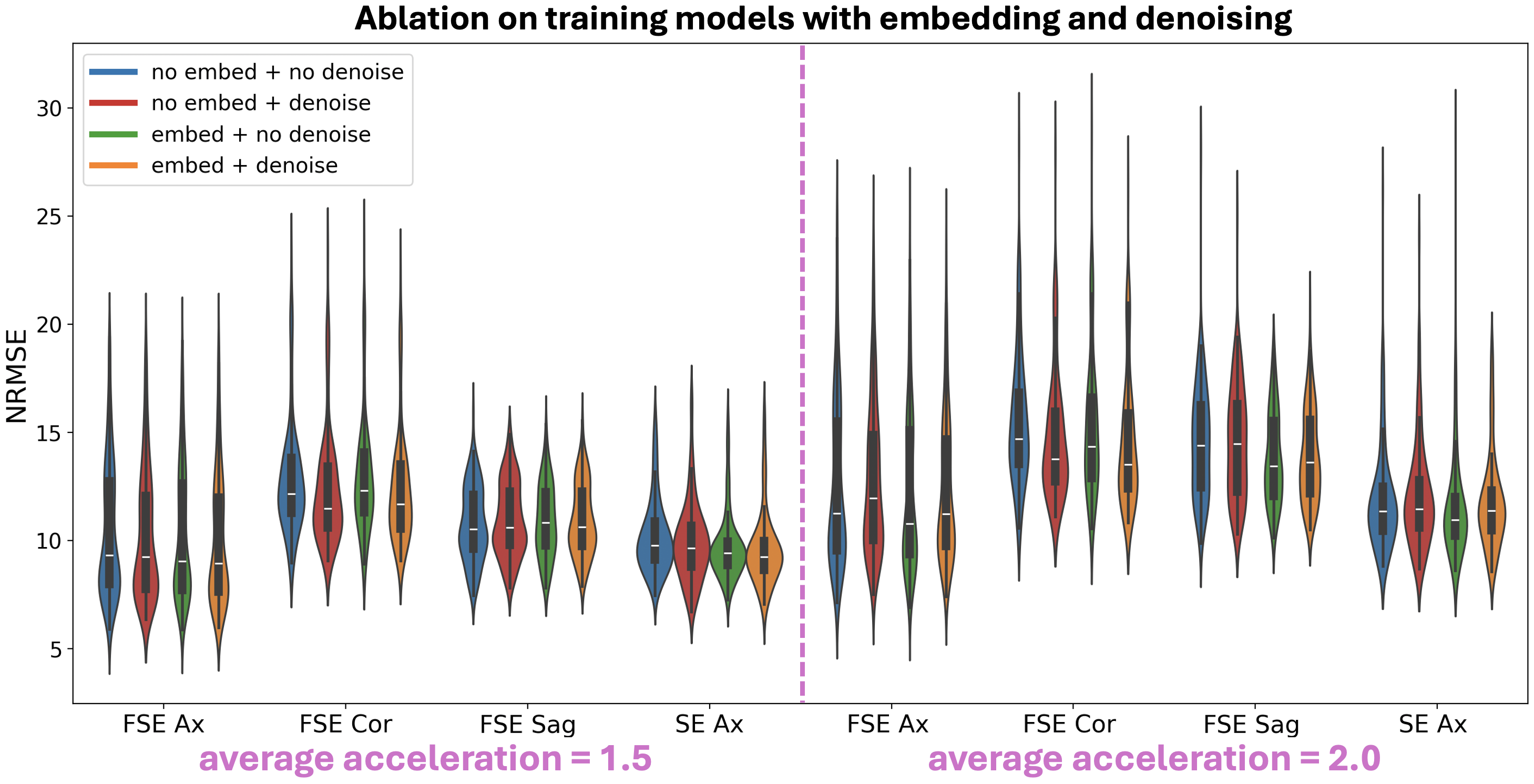}}
    \caption{Violin plots comparing NRMSE across the dataset of posterior sampling reconstructions on 1.5x and 2x under-sampled data using diffusion models trained on all data with an ablation on using class embeddings and denoising. While incorporating class embeddings improves performance, denoising does not always yield quantitative improvement.}
\end{figure*}

\begin{figure*}[t!]
    \centering
    \centerline{\includegraphics[width=\linewidth]{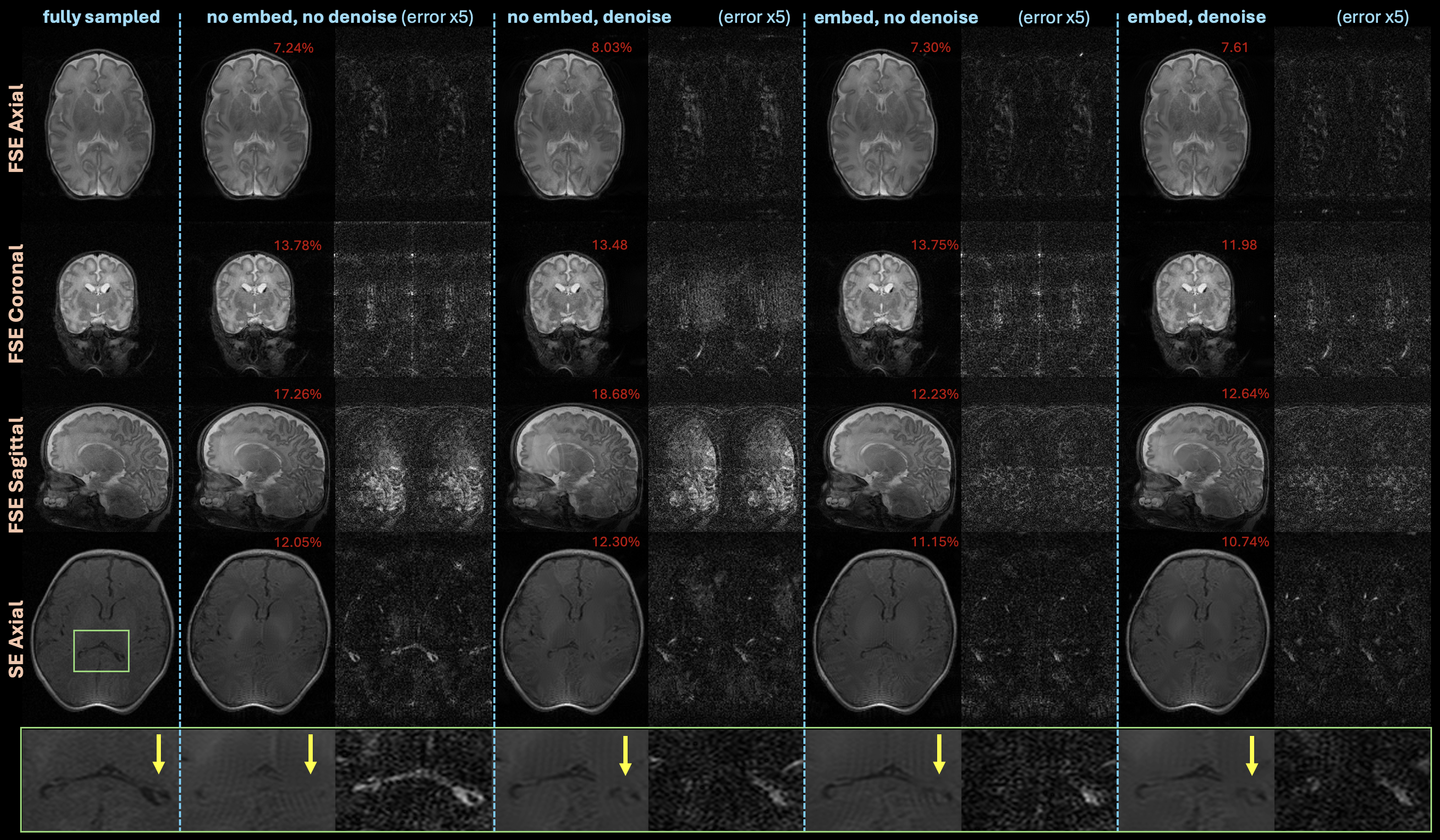}}
    \caption{Example reconstructions from the experiment comparing posterior sampling reconstructions on $R=2$ under-sampled data using diffusion models trained on all data with an ablation on using class embeddings and denoising. In the SE axial results row with zoomed in images, notice how no embedding without denoising yields lower NRMSE than with denoising, but suffers from increased hallucinations.}
\end{figure*}

\subsection{Reader Study}
\subsubsection{Evaluation 1: Blind Comparison of Volumes}
Figure 9 (A) compares the image quality scores from the radiologists when evaluating conventional, fully-sampled and proposed, $1.5 \times$ accelerated volumes in a blind and randomized fashion. For SNR, CNR, artifact level, and overall quality, [conventional, proposed] achieved average scores of \textit{SNR}: [$2.3\pm 0.6, 1.9\pm 0.7$], \textit{CNR}: [$2.6\pm 0.5, 2.3\pm 0.6$], \textit{artifact level}: [$2.2\pm 0.7, 1.8\pm 0.7$], and \textit{overall quality}: [$2.4\pm 0.5, 2.2\pm 0.6$]. A  paired, nonparametric Wilcoxon signed-rank test suggests a statistically significant difference ($p<0.05$) existed between the conventional and proposed volumes for all image quality metrics. Intraclass correlation coefficients (ICC) indicate that there was moderate or strong (ICC $0.53-0.85$) agreement between the two readers for all image quality criteria.

Figure 9 (B) compares the corresponding structure delineation scores. For white matter, cortex, cerebellum, brain stem, and grey matter, [conventional, proposed] yielded average scores of \textit{white matter}: [$2.6\pm 0.5, 2.4\pm 0.5$], \textit{cortex}: [$2.7\pm 0.5, 2.6\pm 0.5$], \textit{cerebellum}: [$2.9\pm 0.3, 2.5\pm 0.5$], \textit{brain stem}: [$3.0\pm 0.2, 2.5\pm 0.5$], and \textit{grey  matter}: [$2.7\pm 0.5, 2.7\pm 0.5$].  Paired nonparametric Wilcoxon tests found statistically significant difference ($p<0.05$) between the conventional and proposed for delineation of white matter, cerebellum, and brain stem but no significant difference for delineation of cortex and grey matter. Strong agreement (ICC $0.71-0.89$) existed between the readers for delineation of all anatomy except cortex, where there was poor agreement (ICC $0.22$).

\begin{figure*}[t!]
    \centering
    \centerline{\includegraphics[width=\linewidth]{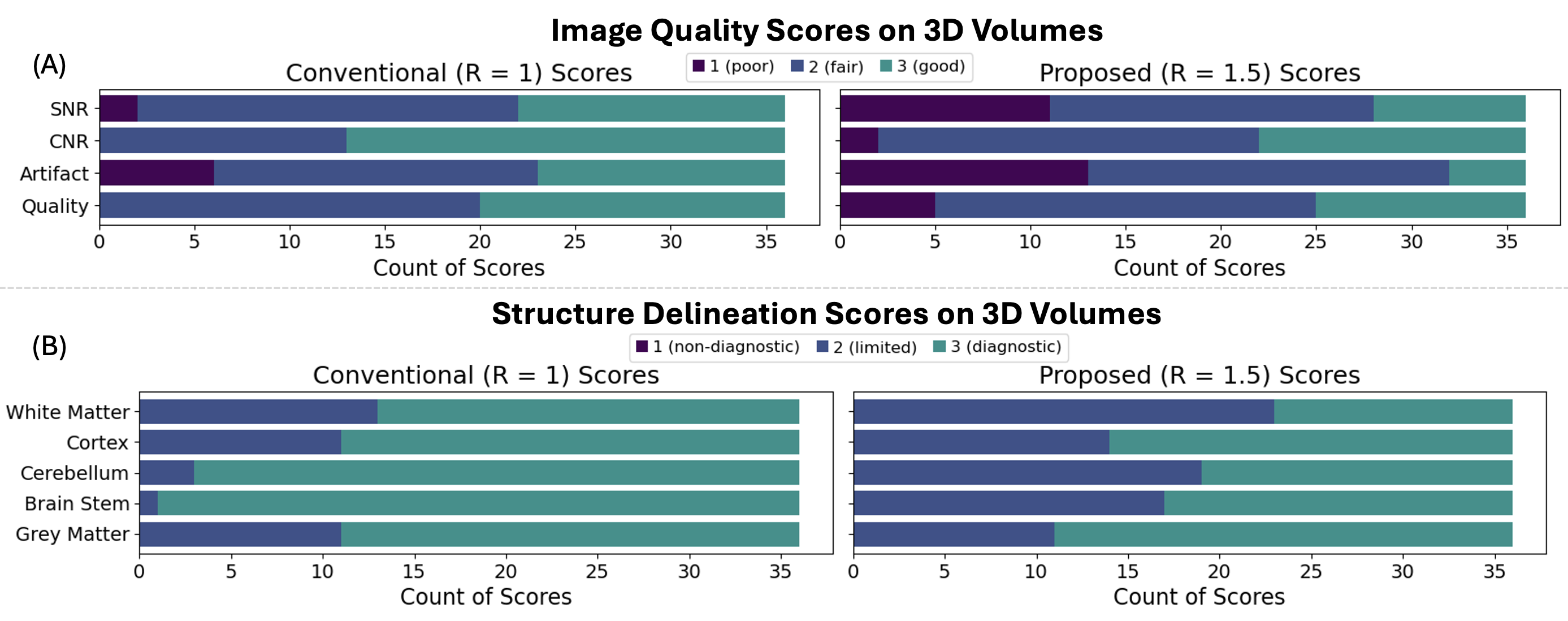}}
    \caption{Count of scores from blind ratings of conventional volumes, reconstructed from fully-sampled data, and proposed volumes, reconstructed from $R=1.5$ under-sampled data, for (A) image quality and (B) structure delineation. While the proposed approach is slightly inferior in the image quality metrics and delineation of white matter, cerebellum, and cortex, it achieved non-inferior delineation of grey matter and cortex. In addition, the proposed approach did not receive any structure delineation scores lower than 2, indicating that all proposed volumes are clinically acceptable while potentially reducing scan time and improving motion robustness.}
\end{figure*}

 Supporting Figure S4 presents the same results from Figure 9 separated by each reader. Supporting Figure S5 shows slices from three pairs of example volumes and their corresponding reader scores. Rows (A) and (B) are two pairs where the accelerated proposed and fully-sampled conventional achieve similar scores, while (C) is a case where the proposed approach rated worse.

\subsubsection{Evaluation 2: Direct Paired Comparison of Slices}
Figure 10 (A) compares the image quality scores from the radiologists when they directly compared pairs of fully-sampled and proposed images. The SNR, CNR, and overall image quality scores were $-0.02\pm 0.8, -0.08\pm 0.8, -0.2\pm 0.9$, and unpaired Wilcoxon signed-rank tests suggests that that the fully-sampled and proposed images are not significantly different with respect to SNR, CNR, and overall image quality (p-values of $0.8, 0.4, 0.06$ respectively). There was fair agreement between the readers (ICC $0.29-0.38$). Figure 10 (B) counts the scores of the two readers for the proposed method adding or removing structure. For removal, $98.6\%$ of the scores suggested that nothing was removed by the proposed approach, while $1.4\%$ indicated that slight structure was removed. For addition, $83.3\%$ of the scores indicate no structure was added, $12.5\%$ suggest slight structure was added, and $2.8\%$ of the scores suggest structure was added. 

\begin{figure*}[t!]
    \centering
    \centerline{\includegraphics[width=\linewidth]{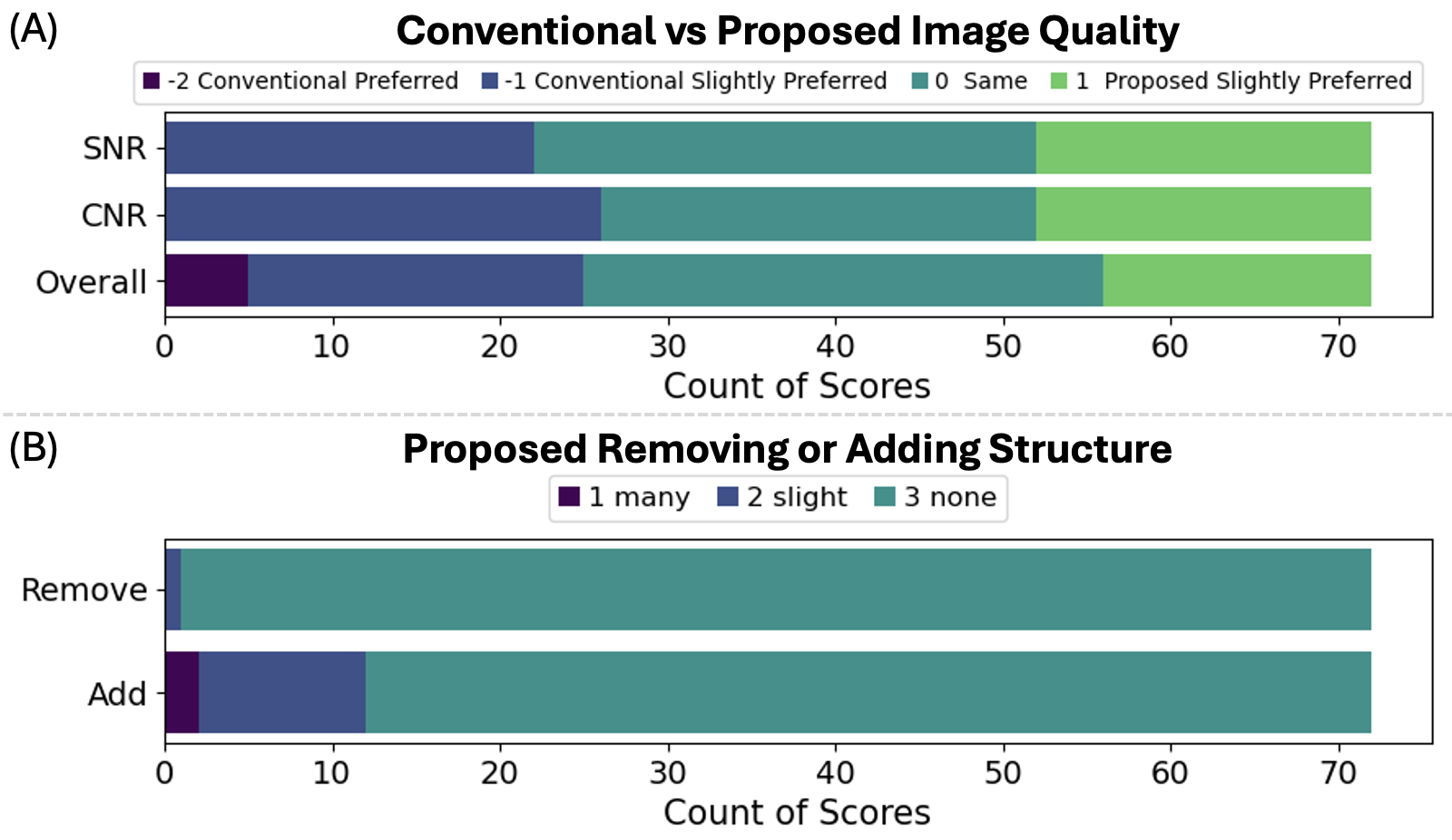}}
    \caption{Count of scores when comparing the proposed approach reconstructed from $R=1.5$ under-sampled data and conventional images reconstructed from fully-sampled data (A) for SNR, CNR, and overall image quality and (B) evaluating whether the proposed approach adds or removes structure. No significant difference was observed between the standard and proposed approach for SNR, CNR, and quality. In addition, the proposed approach mostly did not remove or add much structure, but the most common failure case was the addition of slight structure.}
\end{figure*}

Supporting Figure S6 presents the same results from Figure 10 separated by each reader. Supporting Figure S7 displays three example fully-sampled (no acceleration) and proposed ($R\approx1.5$) pairs and their corresponding reader scores. (A) and (B) show two cases where the proposed approach achieves similar image quality to the fully-sampled image while not missing or adding any structure. (C) illustrates a pair where the readers preferred the fully-sampled image for SNR, CNR, and quality, and they observed some structure hallucinated by the proposed approach.

\section{Discussion}
In this manuscript, we gathered a real-world in-NICU neonatal MRI dataset in collaboration with Aspect Imaging using the 1T Embrace System in Sha'are Zedek Medical Center. Then we propose a generative model training pipeline to address the challenges of limited quantity and low SNR presented by our dataset by combining all available data with class embeddings and denoising the data pre-training.  The clinical reader study suggests that the proposed approach accelerates T$_2$-weighted FSE and T$_1$-weighted spin echo acquisitions by a factor of $1.5\times$ in comparison to the current standard-of care-while maintaining diagnostic utility. Thus even with single coil acquisitions, the proposed framework could reduce total exam times, potentially making MRI more accessible to the most vulnerable and sick patients in the NICU who cannot stay still for longer scan times.

Figures 3 and 7 suggest that combining all data with class embeddings improves reconstruction performance in comparison to training separate models for each contrast and slice orientation. Models trained separately stretches the data too thin while the proposed uses all available training, and the embedding guides posterior sampling during inference. Since contrast and orientation is known apriori in MRI, the embedding is known during reconstruction. Training the model on all of the data also could improve performance on out of distribution measurements \cite{lin2023robustness}. Future work will explore additional metadata context \cite{chung2025context}. To further address limited data, the model could be pre-trained on open adult MRI datasets and then fine-tuned using our real-world, clinical in-NICU dataset \cite{hu2022lora, kumar2025ismrm}.

On the other hand, Figures 5 and 7 do not always show clear improvements when using denoising as a pre-training step. Our quantitative metrics are computed with respect to fully-sampled, but still inherently noisy images without denoising, so this may bias comparisons between our proposed method with and without denoising pre-training \cite{haldarnoise}. The SE axial result in Figure 8 also suggests that quantitative comparisons can be deceptive \cite{chan2023ReferenceBasedIQA}, where the no embed, no denoise model achieves lower NRMSE than the no embed, denoising model, but the zoomed in view shows that the lower NRMSE approach misses more anatomy. We performed our reader study to address this limitation in quantitative comparisons. 

Noise variances estimated from the $20 \times 20$ background patches were used for self-supervised training of the denoiser. While estimating noise statistics from background patches creates vulnerability to bias from effects like stimulated echoes, aliasing, and gradient nonlinearities, Supporting Figure S8 plots histograms of signal in the background patch of $25$ random training samples with the associated gaussisan fit. Qualitatively, the signal appears similar to zero-mean, gaussian noise, but acquisition of any future training data will keep the noise pre-scan to estimate noise variance.

Figure 6 shows that averaging posterior samples monotonically improves quantitative performance, but with diminishing returns, similar to adult MRI \cite{luo2023diff}. This work heuristically selected five posterior samples to estimate the conditional expectation to balance computation time with performance, but different applications might require more or fewer samples. Fast reconstruction times are important for clinical adoption, but diffusion models are computationally intensive \cite{jalal2021robust,chung2023dps}. Our parallel implementation reconstructs the five posterior samples for averaging in approximately $17$ seconds on a NVIDIA H100 GPU, and future efficient implementations will further parallelize along the slice dimension for fast volumetric reconstructions (current memory usage allows 6 parallel slices).

Previous theoretical results show that posterior sampling is close to the optimal algorithm for reconstruction from an arbitrary measurement process \cite{jalal2021robust}. In practice, we approximate the prior distribution through training a neural network on a noisy, finite training dataset using denoising score-matching, and we approximate the posterior sampling algorithm itself with Diffusion Posterior Sampling \cite{chung2023dps}. As a result, there is a known theoretical gap between the computational approach and convergence guarantees \cite{gupta2024psintract}. Nonetheless, we demonstrate the effectiveness of our framework with empirical results on real-world in-NICU neonatal MRI, though future work is required to bridge the theoretical gap.

We selected diffusion based generative models in this work for their robustness to changes in the measurement model. Figure 7 shows that the same generative model applies to two different measurement models with under-sampling rates of $R = 1.5$ and $R = 2.0$ without any modifications to posterior sampling hyperparameters. This flexibility also extends to retrospective motion correction where the measurement model might change for each test sample \cite{frost2022mocoreview,zaitsev2016motionreview, levac2024moco}. In Supporting Figure S9, we use the algorithm detailed in Reference \cite{levac2024moco} to apply retrospective motion correction to prospectively acquired, motion corrupt in-NICU neonatal data. This preliminary example shows that combining generative models with motion-based measurement models could reduce image artifacts. In addition, since the generative prior decouples from the measurement model, it applies to motion patterns without requiring that the training data contain examples of motion. Future work will involve further technical development and experimental validation exploring the application of generative models for retrospective motion correction of in-NICU neonatal MRI.

Since previous literature extensively compares generative models to other state-of-the-art methods for accelerated MRI reconstruction \cite{jalal2021robust,chung2022score,luo2023diff}, this manuscript focused on analyzing our contributions in developing a novel pipeline for training generative models on our in-NICU neonatal MRI data and performing the first clinical validation of generative models for accelerated in-NICU neonatal MRI. However, for the interested reader, Supporting Figures S1 and S2 show that generative models trained with the proposed framework achieve competitive quantitative reconstruction performance in comparison to MoDL, a state-of-the-art end-to-end method. We suspect ModL performed worse on the axial images because the test dataset consisted of different under-sampling patterns and matrix sizes from the training data (since we under-sampled by throwing away groups of echo trains which varied across the test set), which ModL perhaps did not adapt well too. Future research directions could develop methods to most effectively train end-to-end methods for accelerated in-NICU neonatal MRI. Note, generative methods provide the additional benefit of applying to other measurement models without re-training, like motion correction described in the previous paragraph.

Other ``plug-and-play'' methods for accelerated MRI reconstruction also exist that decouple the prior from the likelihood \cite{ahmad2020pnpmri, reehorst2018redclarity, chand2024energy, liu2020rare}. In addition, some of these methods do not require averaging of posterior samples for image reconstruction. We selected generative models as a starting point for in-NICU neonatal MRI based on previous literature that found generative models achieve competitive performance in comparison to other state-of-the-art learning based accelerated MRI reconstruction techniques while decoupling the prior and likelihood \cite{jalal2021robust,chung2022score,luo2023diff,zheng2025inversebench}. Thus, we focused our contribution on proposing a pipeline to effectively train generative models in the challenging in-NICU setting and thoroughly evaluate the method with our clinical reader study. In addition, generative models enable potential estimates of reconstruction uncertainty by computing statistics, such as standard deviation, over the posterior samples. For example, Supporting Figure S10 shows how standard deviation across the posterior samples from Figure 6 increases in regions of higher reconstruction error. Future work will evaluate other plug-and-play methods in the in-NICU neonatal MRI setting.

The reader study evaluated conventional fully-sampled images and proposed images reconstructed from $R=1.5$ under-sampled data. The blind evaluation of volume scores from Figure 9 (A) indicate that proposed volumes are inferior to the conventional for SNR, CNR, artifact level, and overall image quality. However, the maximum difference in any one of the metrics was $13\%$, suggesting that the proposed approach is just slightly worse. In addition, when the readers directly compared paired conventional and proposed slices, illustrated in Figure 10 (A), no significant difference was observed in SNR, CNR, and overall image quality. 

Since image quality scores do not necessarily correlate with diagnostic utility, the reader study also evaluated structure delineation. The blind volumetric comparison found that the proposed approach was non-inferior to conventional images for cortex and brain stem delineation, but inferior for delineation of white matter, cerebellum, and brain stem. However, the average scores for white matter, cerebellum, and brain stem only differed by $7\%$, $13\%$, and $16\%$ respectively. In addition when discussing the scoring system, the radiologists explained that a score of 2 (limited) for structure delineation implies that an image would be clinically acceptable to evaluate that anatomy. Our proposed approach received an average score of $2.3$ and higher for delineation of all structure in the study and did not receive any scores less than 2, indicating that all proposed images in this study would be useful clinically while potentially reducing scan time by $1.5\times$ and reducing motion sensitivity.

Since our method relies on the generative model to handle the ill-posed under-sampled reconstruction problem, the final phase of the reader study evaluated whether the proposed approach hallucinated or removed structure present in the fully-samples images. Figure 10 (B) shows that the model rarely removes structure, but adds slight structure in a small percentage of cases. An additional evaluation will explore whether these slight hallucinations hinder diagnostic accuracy and if further technical development for mitigating hallucinations is required.

Although we considered signal evolution to make under-sampling as realistic as possible \cite{FSE_comp_moco,rajput2024retro}, all experiments still utilized retrospective under-sampling. In addition, while structure delineation is important for diagnosis, our study did not compare diagnostic outcomes from using fully-sampled versus proposed images. A future reader study, with images reconstructed from prospectively under-sampled data, will compare radiologists' diagnoses made with the full suite of contrasts used clinically.

\section{Conclusion}
This work proposed accelerating in-NICU neonatal MRI with diffusion probabilistic generative models. We gathered a clinical dataset with Aspect Imaging and Sha'are Zedek Medical Center, and trained the diffusion models by combining all heterogeneous data with class embeddings and applying self-supervised denoising to our dataset. Averaging five diffusion posterior samples reconstructs our desired image. A clinical reader study suggests that our proposed approach could reduce scan time by a factor of $1.5\times$ in fast-spin-echo and spin-echo scans while maintaining acceptable structural delineation of anatomy.

\section{Acknowledgment}
This work was supported in part by Aspect Imaging, NSF CCF-2239687 (CAREER), NSF IFML 2019844, and JCCO fellowship.

\bibliography{references}

\section*{Figure Captions}
\noindent \textbf{Figure 1:} The proposed training and inference pipeline. Images from different orientations and contrasts are combined into a single dataset using class embeddings. A denoiser, trained in a self-supervised fashion, denoises the dataset before training. Diffusion posterior sampling reconstructs an image from under-sampled k-space by averaging 5 posterior samples. The proposed methods enable training of diffusion models for accelerated reconstruction on noisy and limited in-NICU neonatal MRI data. 

\noindent \textbf{Figure 2:} (a) The top row shows two training samples from our dataset and the bottom row shows the corresponding training samples after applying our denoiser trained in a self-supervised fashion. (b,c,d,e) prior samples generated by our trained model when conditioned on class embeddings of fse axial, sagittal, coronal, and se axial. Our model uses all available training data to learn a statistical prior over neonatal mr images.

\noindent \textbf{Figure 3:} Violin plots comparing NRMSE of posterior sampling reconstructions on $R=2$ under-sampled data using diffusion models trained on just a single image class, trained on all data, and trained on all data with class embeddings. Results on all contrasts and orientations suggest that a diffusion model trained by combining all data with class embeddings yields the best quantitative performance. 

\noindent \textbf{Figure 4:} Example images from the experiment comparing posterior sampling reconstructions on $R = 2$ under-sampled data using diffusion models trained on just a single image class, trained on all data, and trained on all data with class embeddings. This is a specific illustration of the quantitative conclusion that using all data combined with class embeddings to train the diffusion model yields best performance.

\noindent \textbf{Figure 5:} Violin plots comparing NRMSE of reconstructions on $R=1.5$ under-sampled data using baseline L$_1$ and diffusion models trained on all data with and without denoising. While using a learned prior provides benefit, little quantitative difference exists between models trained with and without denoising pre-training.

\noindent \textbf{Figure 6:} (A) Mean NRMSE of a generative model trained with class embeddings on all data with pre-training denoising when averaging \{$1,2,3,4,5$\} posterior samples. Averaging more posterior sample leads to improved quantitative performance. (B) Two specific examples from the experiment where averaging more posterior samples improves quantitative and qualitative performance.

\noindent \textbf{Figure 7:} Violin plots comparing NRMSE across the dataset of posterior sampling reconstructions on 1.5x and 2x under-sampled data using diffusion models trained on all data with an ablation on using class embeddings and denoising. While incorporating class embeddings improves performance, denoising does not always yield quantitative improvement.

\noindent \textbf{Figure 8:} Example reconstructions from the experiment comparing posterior sampling reconstructions on $R=2$ under-sampled data using diffusion models trained on all data with an ablation on using class embeddings and denoising. In the SE axial results row with zoomed in images, notice how no embedding without denoising yields lower NRMSE than with denoising, but suffers from increased hallucinations.

\noindent \textbf{Figure 9:} Count of scores from blind ratings of conventional volumes, reconstructed from fully-sampled data, and proposed volumes, reconstructed from $R=1.5$ under-sampled data, for (A) image quality and (B) structure delineation. While the proposed approach is slightly inferior in the image quality metrics and delineation of white matter, cerebellum, and cortex, it achieved non-inferior delineation of grey matter and cortex. In addition, the proposed approach did not receive any structure delineation scores lower than 2, indicating that all proposed volumes are clinically acceptable while potentially reducing scan time and improving motion robustness.

\noindent \textbf{Figure 10:} Count of scores when comparing the proposed approach reconstructed from $R=1.5$ under-sampled data and conventional images reconstructed from fully-sampled data (A) for SNR, CNR, and overall image quality and (B) evaluating whether the proposed approach adds or removes structure. No significant difference was observed between the standard and proposed approach for SNR, CNR, and quality. In addition, the proposed approach mostly did not remove or add much structure, but the most common failure case was the addition of slight structure.

\section*{Supporting Figure Captions}
\noindent \textbf{Supporting Figure S1:} Quantitative NRMSE results on all four image classes comparing reconstructions on $R=2$ under-sampled data using a Diffusion model trained on just the image class, ModL trained on just the image class, Diffusion trained on all classes, MoDL trained on all classes, and the proposed diffusion framework trained on all data with class embeddings.

\noindent \textbf{Supporting Figure S2:} Examples reconstructions on all four image classes comparing reconstructions on $R=2$ under-sampled data using a Diffusion model trained on just the image class, ModL trained on just the image class, Diffusion trained on all classes, MoDL trained on all classes, and the proposed diffusion framework trained on all data with class embeddings.

\noindent \textbf{Supporting Figure S3:} Examples from the experiment comparing reconstructions on $R=1.5$ under-sampled data using an L$_1$ baseline and diffusion models trained on all data with and without denoising. A learned prior provides benefit, but models with and without denoising do not exhibit clear differences in performance.

\noindent \textbf{Supporting Figure S4:} Count of scores from blind ratings of conventional volumes where the scores have been split by reader (R1 and R2). The proposed approach did not receive any structure delineation scores lower than 2, indicating that all volumes reconstructed from $R=1.5$ under-sampled data are clinically acceptable.

\noindent \textbf{Supporting Figure S5:} Example slices from volumes and their corresponding scores for the first evaluation in the reader study in which volumes were compared in a blinded fashion. (A) and (B) illustrate cases where the proposed approach matches the conventional images in quality, while (C) shows an example where the proposed approach was rated worse than the conventional.

\noindent \textbf{Supporting Figure S6:} Count of scores, separated by readers (R1 and R2) when directly comparing proposed and conventional image slices for image quality and whether structure was added or removed. No difference was observed in image quality and the proposed mostly did not remove or add structure, but the most common failure case was the addition of slight structure.

\noindent \textbf{Supporting Figure S7:} Example image pairs and their corresponding scores from the reader study. (A) and (B) illustrate cases where the proposed approach, reconstructed from $R=1.5$ under-sampled data, maintains similar image quality to the conventional image. (C) shows an example where the proposed approach hallucinated slight additional structure and rated worse for image quality.

\noindent \textbf{Supporting Figure S8:} Histograms of signal in the background $20 \times 20$ patches with the associated gaussian fit in dashed blue of 25 random training samples used to estimate noise variance for self-supervised training of the denoiser. Qualitatively, the signal appears similar to zero-mean, gaussian noise.  

\noindent \textbf{Supporting Figure S9:} An example of applying our generative model to the task of retrospective motion correction where the inverse problem simultaneously estimates the motion parameters and associated image from motion corrupt data. Since generative models decouple the image prior and measurement model, our proposed generative model readily applies to this task of motion correction without any re-training. Future work will involve further technical development and experimental validation exploring the application of generative models for retrospective motion correction of in-NICU neonatal MRI. 

\noindent \textbf{Supporting Figure S10:} Example FSE Sagittal and SE Axial fully-sampled images and associated reconstruction from one posterior sample. The corresponding error and standard deviation across multiple posterior samples show that reconstruction with posterior sampling using generative models enables potential estimates of reconstruction uncertainty by computing statistics over multiple posterior samples.

\newpage

\section*{Supporting Information}
\subsection*{Supporting Figure S1}
\renewcommand{\thefigure}{S1}
\begin{figure}[H]
    \centering
    \centerline{\includegraphics[width=\linewidth]{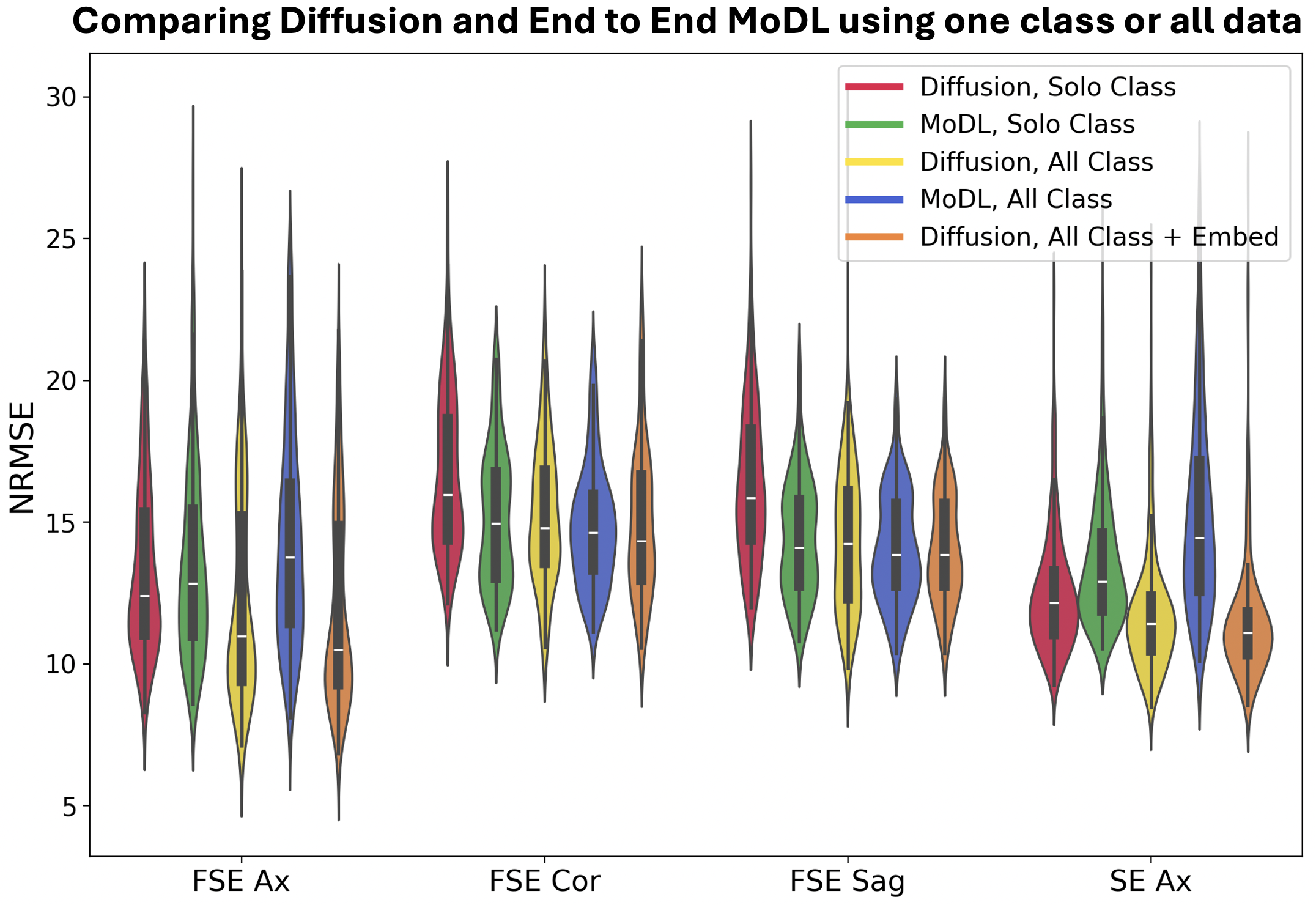}}
    \caption{Quantitative NRMSE results on all four image classes comparing reconstructions on $R=2$ under-sampled data using a Diffusion model trained on just the image class, ModL trained on just the image class, Diffusion trained on all classes, MoDL trained on all classes, and the proposed diffusion framework trained on all data with class embeddings.}
\end{figure}
\renewcommand{\thefigure}{\arabic{figure}}  

\newpage

\subsection*{Supporting Figure S2}
\renewcommand{\thefigure}{S2}
\begin{figure}[H]
    \centering
    \centerline{\includegraphics[width=\linewidth]{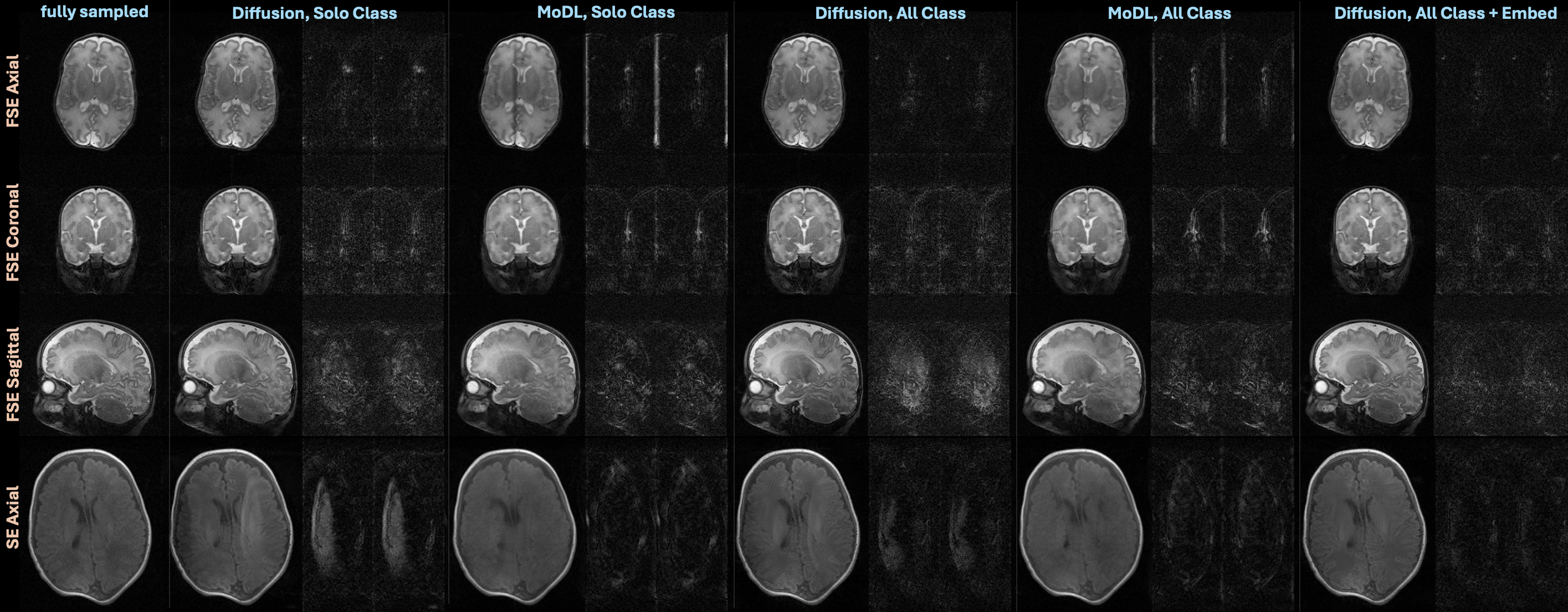}}
    \caption{Examples reconstructions on all four image classes comparing reconstructions on $R=2$ under-sampled data using a Diffusion model trained on just the image class, ModL trained on just the image class, Diffusion trained on all classes, MoDL trained on all classes, and the proposed diffusion framework trained on all data with class embeddings.}
\end{figure}
\renewcommand{\thefigure}{\arabic{figure}}  

\newpage

\subsection*{Supporting Figure S3}
\renewcommand{\thefigure}{S3}
\begin{figure}[H]
    \centering
    \centerline{\includegraphics[width=\linewidth]{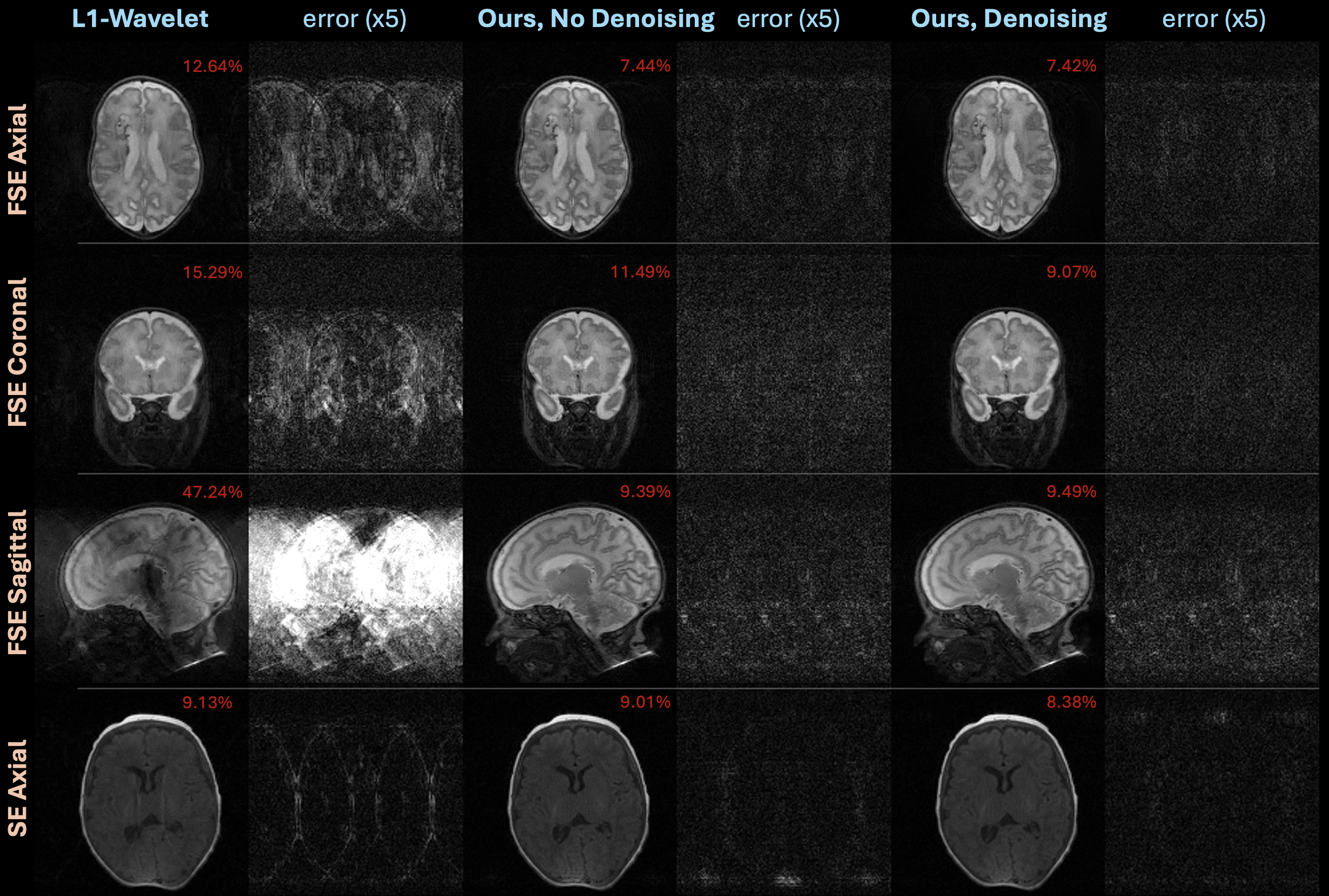}}
    \caption{Examples from the experiment comparing reconstructions on $R=1.5$ under-sampled data using an L$_1$ baseline and diffusion models trained on all data with and without denoising. A learned prior provides benefit, but models with and without denoising do not exhibit clear differences in performance.}
\end{figure}
\renewcommand{\thefigure}{\arabic{figure}}  

\newpage

\subsection*{Supporting Figure S4}
\renewcommand{\thefigure}{S4}
\begin{figure}[H]
    \centering
    \centerline{\includegraphics[width=\linewidth]{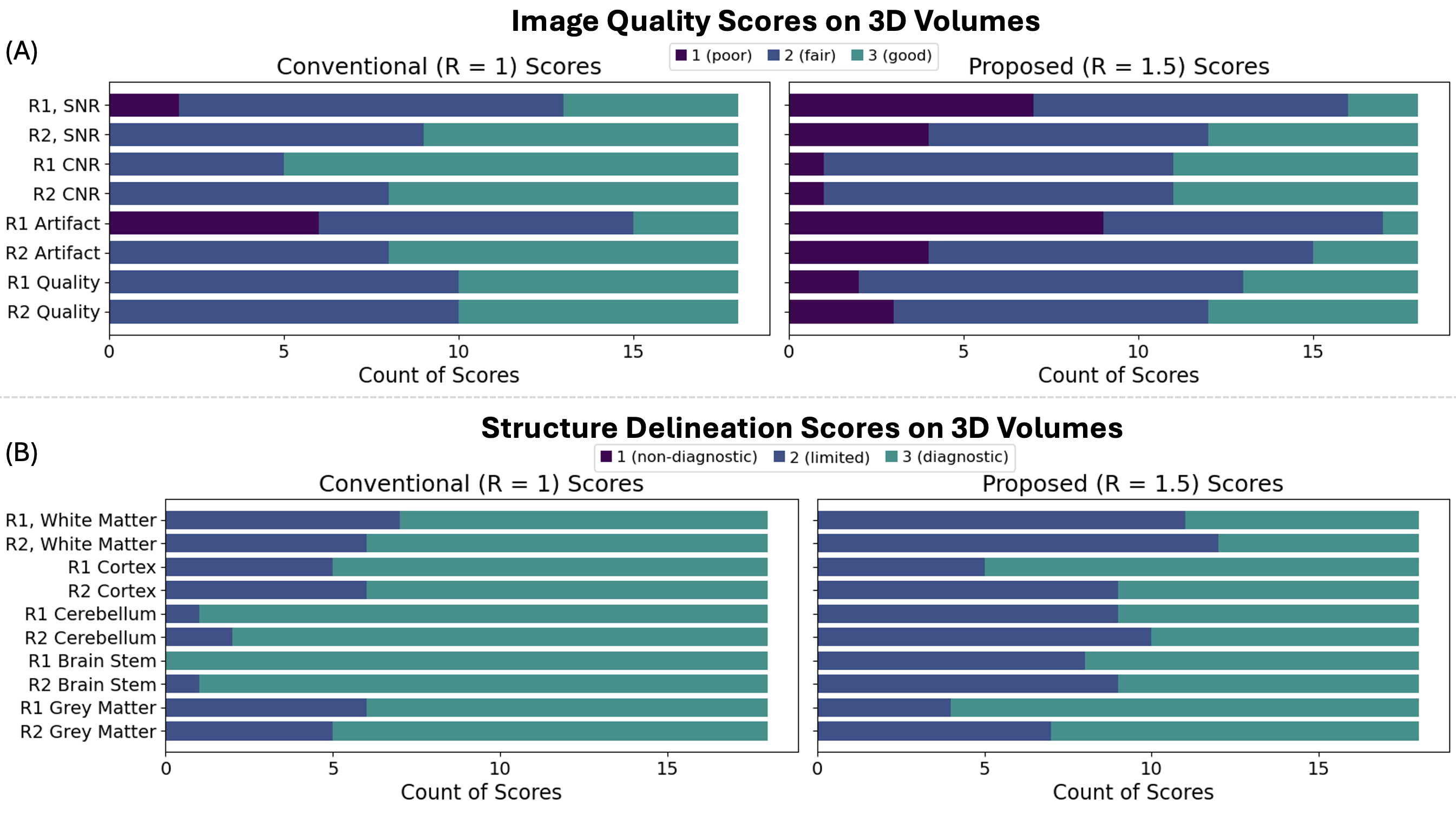}}
    \caption{Count of scores from blind ratings of conventional volumes where the scores have been split by reader (R1 and R2). The proposed approach did not receive any structure delineation scores lower than 2, indicating that all volumes reconstructed from $R=1.5$ under-sampled data are clinically acceptable.}
\end{figure}
\renewcommand{\thefigure}{\arabic{figure}}  

\newpage

\subsection*{Supporting Figure S5}
\renewcommand{\thefigure}{S5}
\begin{figure}[H]
    \centering
    \centerline{\includegraphics[width=\linewidth]{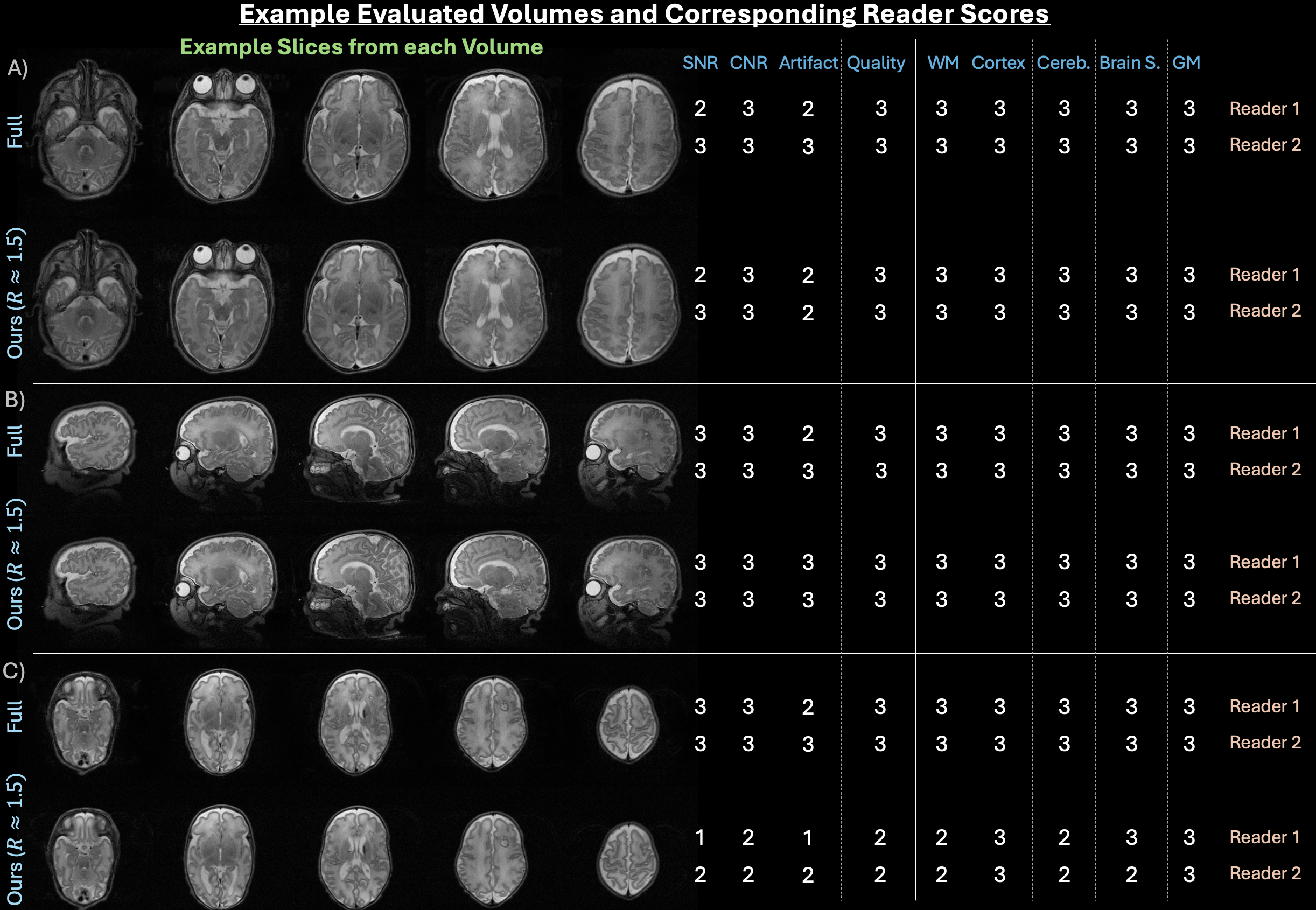}}
    \caption{Example slices from volumes and their corresponding scores for the first evaluation in the reader study in which volumes were compared in a blinded fashion. (A) and (B) illustrate cases where the proposed approach matches the conventional images in quality, while (C) shows an example where the proposed approach was rated worse than the conventional.}
\end{figure}
\renewcommand{\thefigure}{\arabic{figure}}  

\newpage

\subsection*{Supporting Figure S6}
\renewcommand{\thefigure}{S6}
\begin{figure}[H]
    \centering
    \centerline{\includegraphics[width=\linewidth]{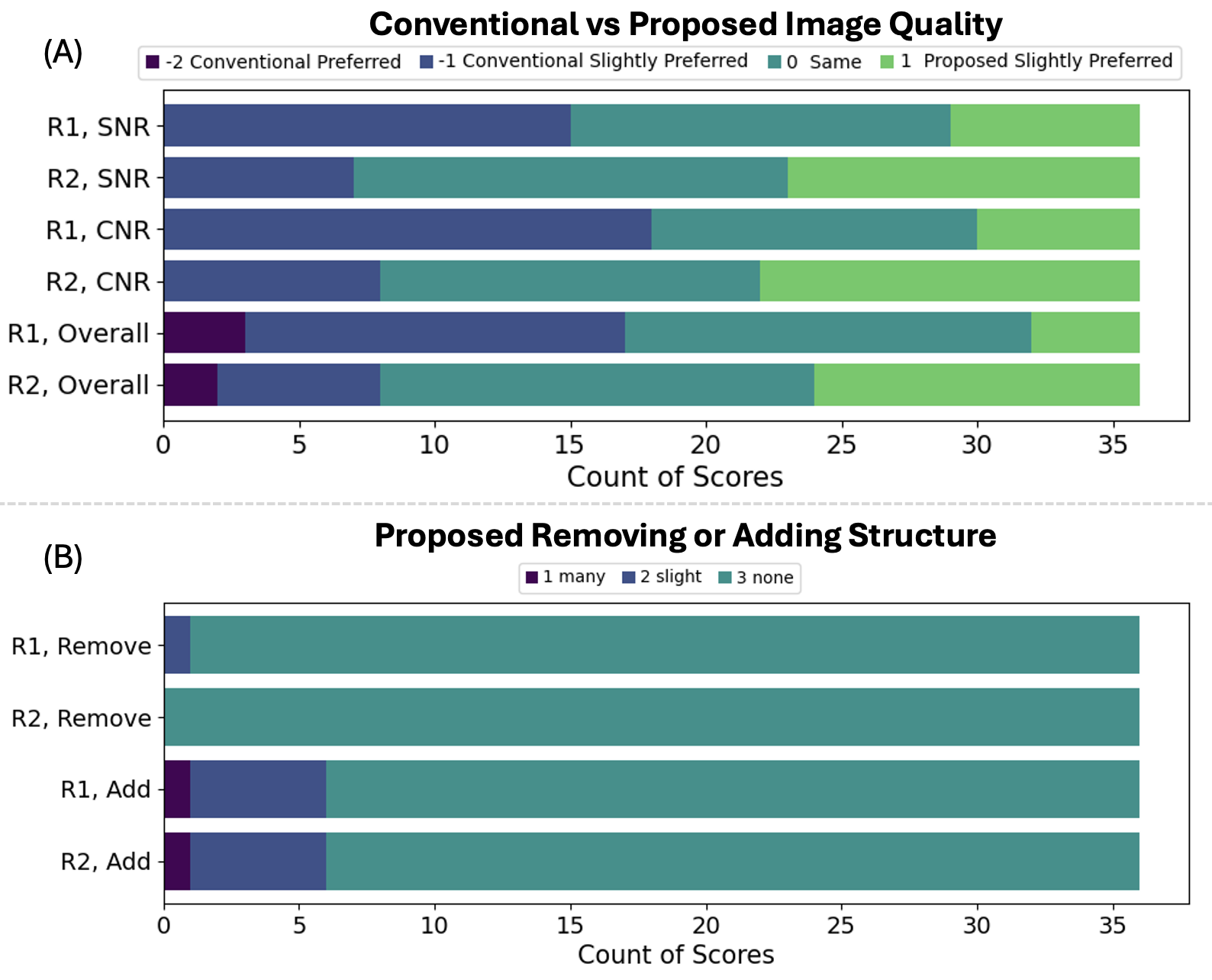}}
    \caption{Count of scores, separated by readers (R1 and R2) when directly comparing proposed and conventional image slices for image quality and whether structure was added or removed. No difference was observed in image quality and the proposed mostly did not remove or add structure, but the most common failure case was the addition of slight structure.}
\end{figure}
\renewcommand{\thefigure}{\arabic{figure}}  

\newpage

\subsection*{Supporting Figure S7}
\renewcommand{\thefigure}{S7}
\begin{figure}[H]
    \centering
    \centerline{\includegraphics[width=\linewidth]{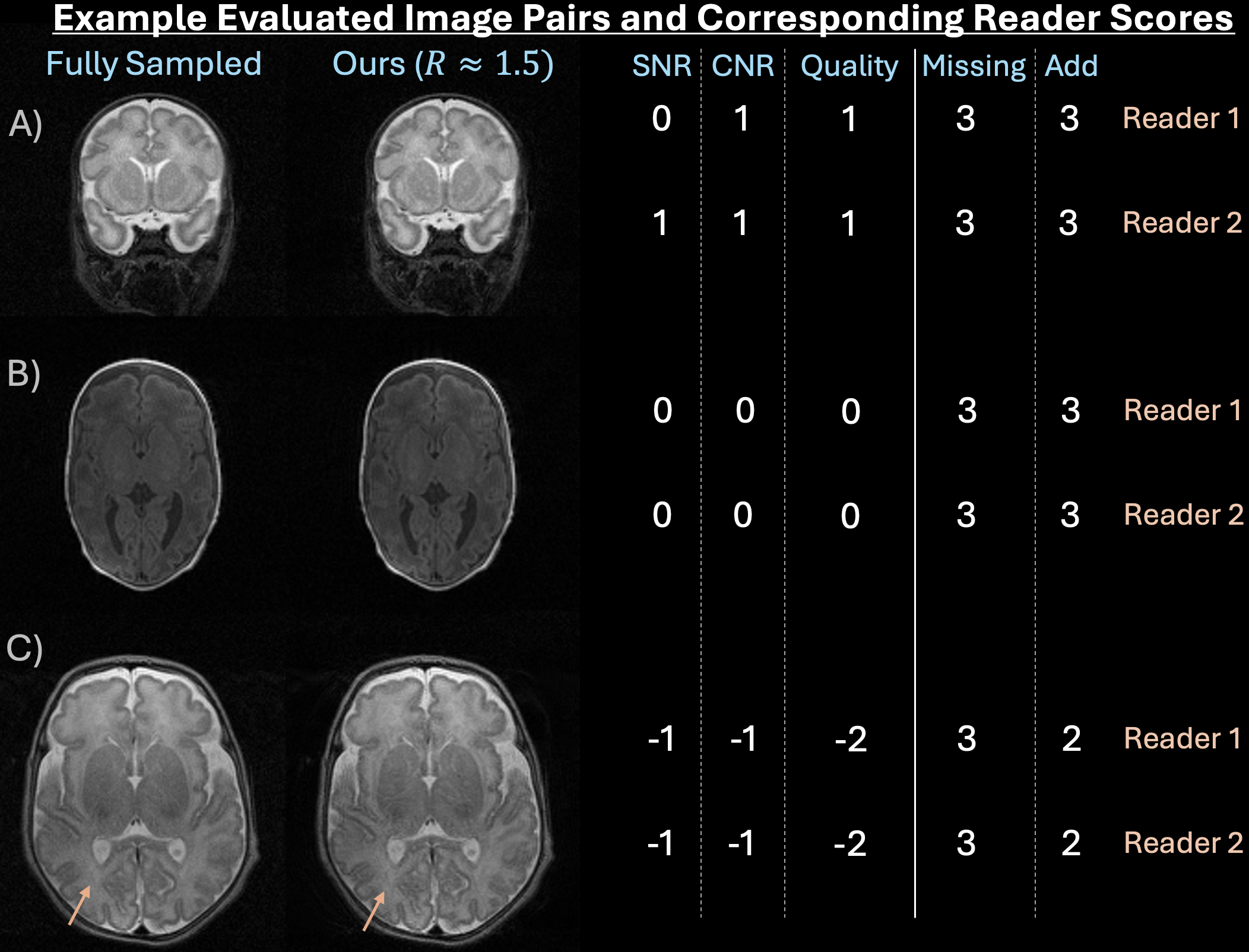}}
    \caption{Example image pairs and their corresponding scores from the reader study. (A) and (B) illustrate cases where the proposed approach, reconstructed from $R=1.5$ under-sampled data, maintains similar image quality to the conventional image. (C) shows an example where the proposed approach hallucinated slight additional structure and rated worse for image quality.}
\end{figure}
\renewcommand{\thefigure}{\arabic{figure}}  

\newpage

\subsection*{Supporting Figure S8}
\renewcommand{\thefigure}{S8}
\begin{figure}[H]
    \centering
    \centerline{\includegraphics[width=\linewidth]{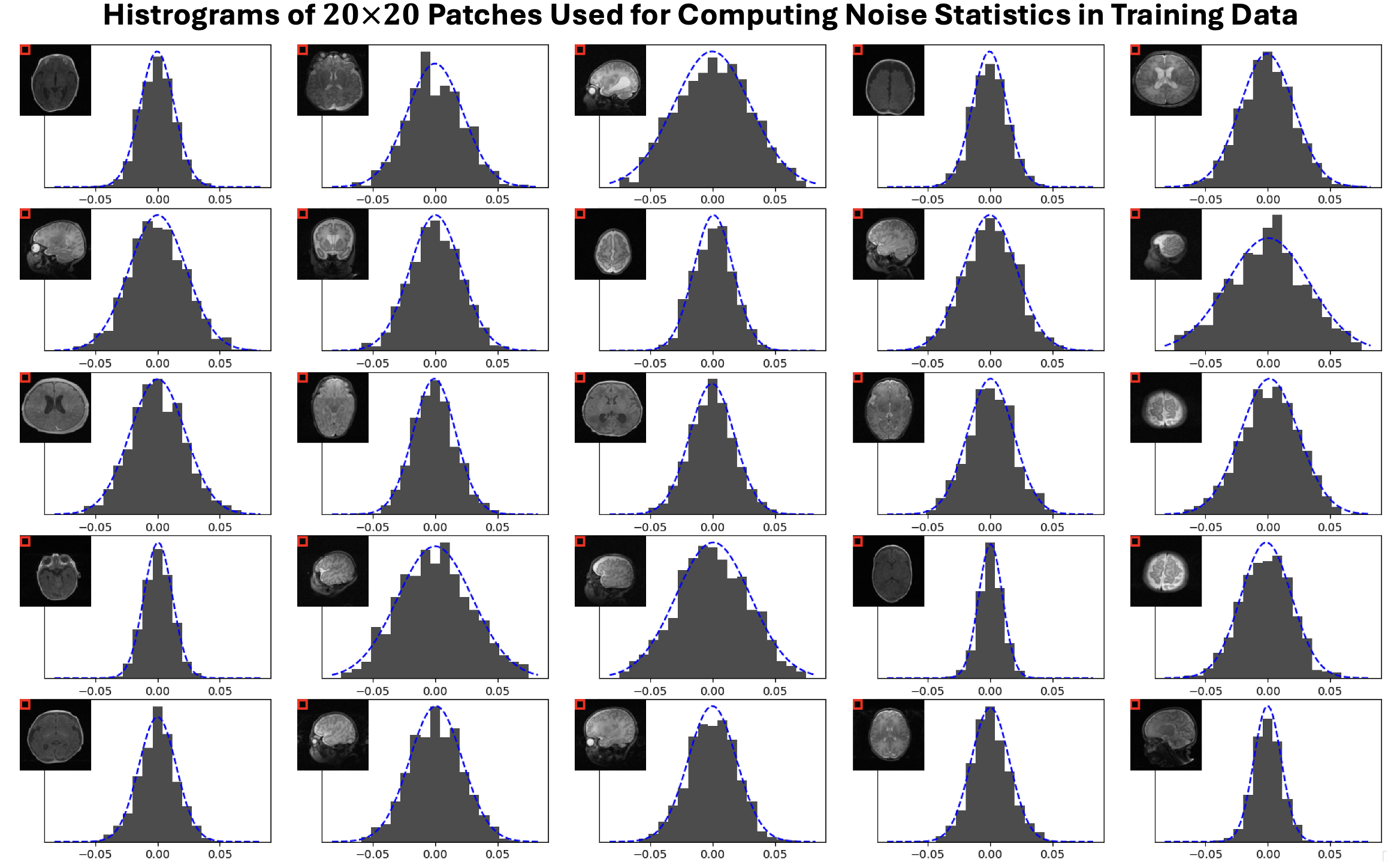}}
    \caption{Histograms of signal in the background $20 \times 20$ patches with the associated gaussian fit in dashed blue of 25 random training samples used to estimate noise variance for self-supervised training of the denoiser. Qualitatively, the signal appears similar to zero-mean, gaussian noise.}
\end{figure}
\renewcommand{\thefigure}{\arabic{figure}}  

\newpage

\subsection*{Supporting Figure S9}
\renewcommand{\thefigure}{S9}
\begin{figure}[H]
    \centering
    \centerline{\includegraphics[width=\linewidth]{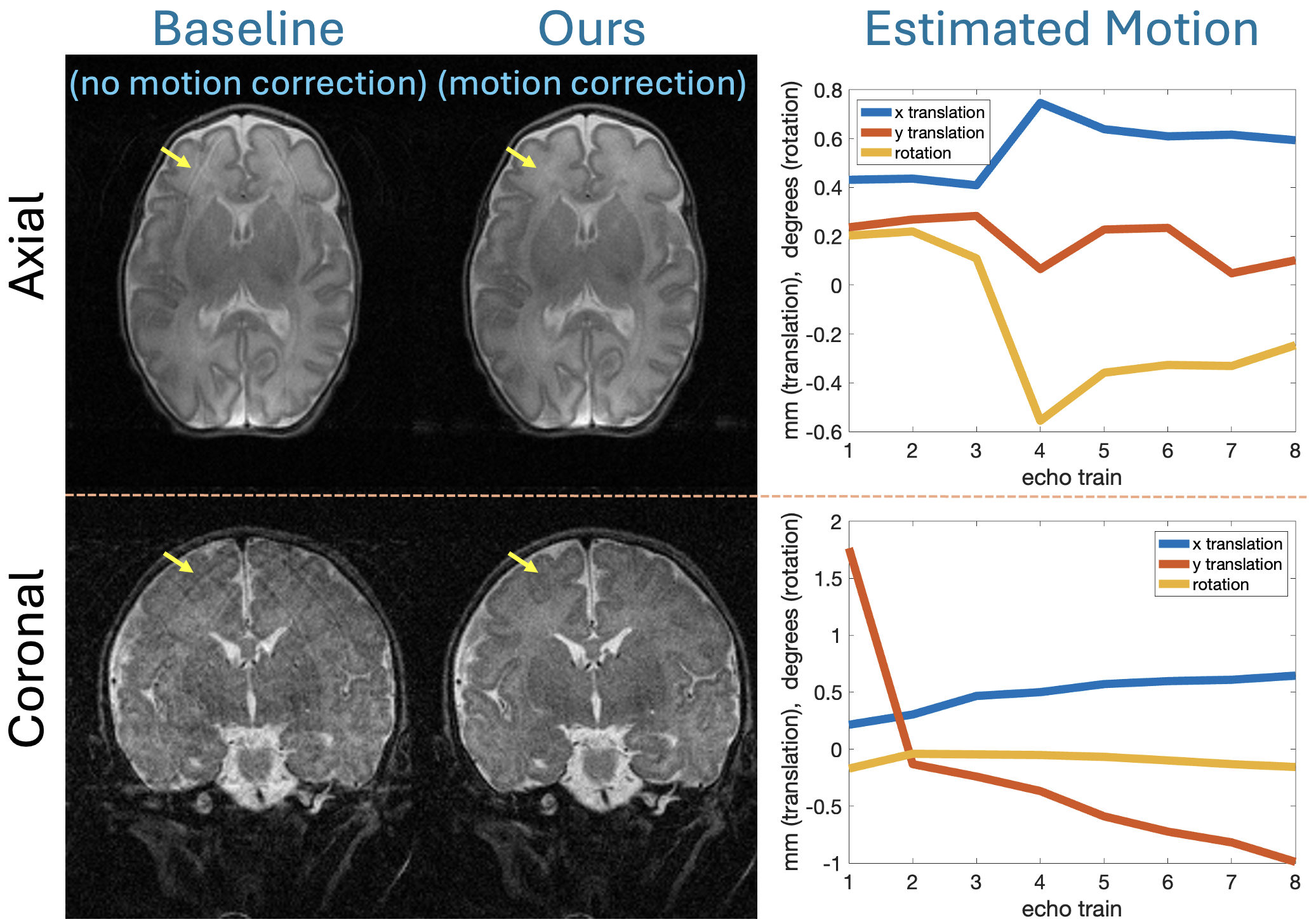}}
    \caption{An example of applying our generative model to the task of retrospective motion correction where the inverse problem simultaneously estimates the motion parameters and associated image from motion corrupt data. Since generative models decouple the image prior and measurement model, our proposed generative model readily applies to this task of motion correction without any re-training. Future work will involve further technical development and experimental validation exploring the application of generative models for retrospective motion correction of in-NICU neonatal MRI.}
\end{figure}
\renewcommand{\thefigure}{\arabic{figure}}  

\newpage

\subsection*{Supporting Figure S10}
\renewcommand{\thefigure}{S10}
\begin{figure}[H]
    \centering
    \centerline{\includegraphics[width=\linewidth]{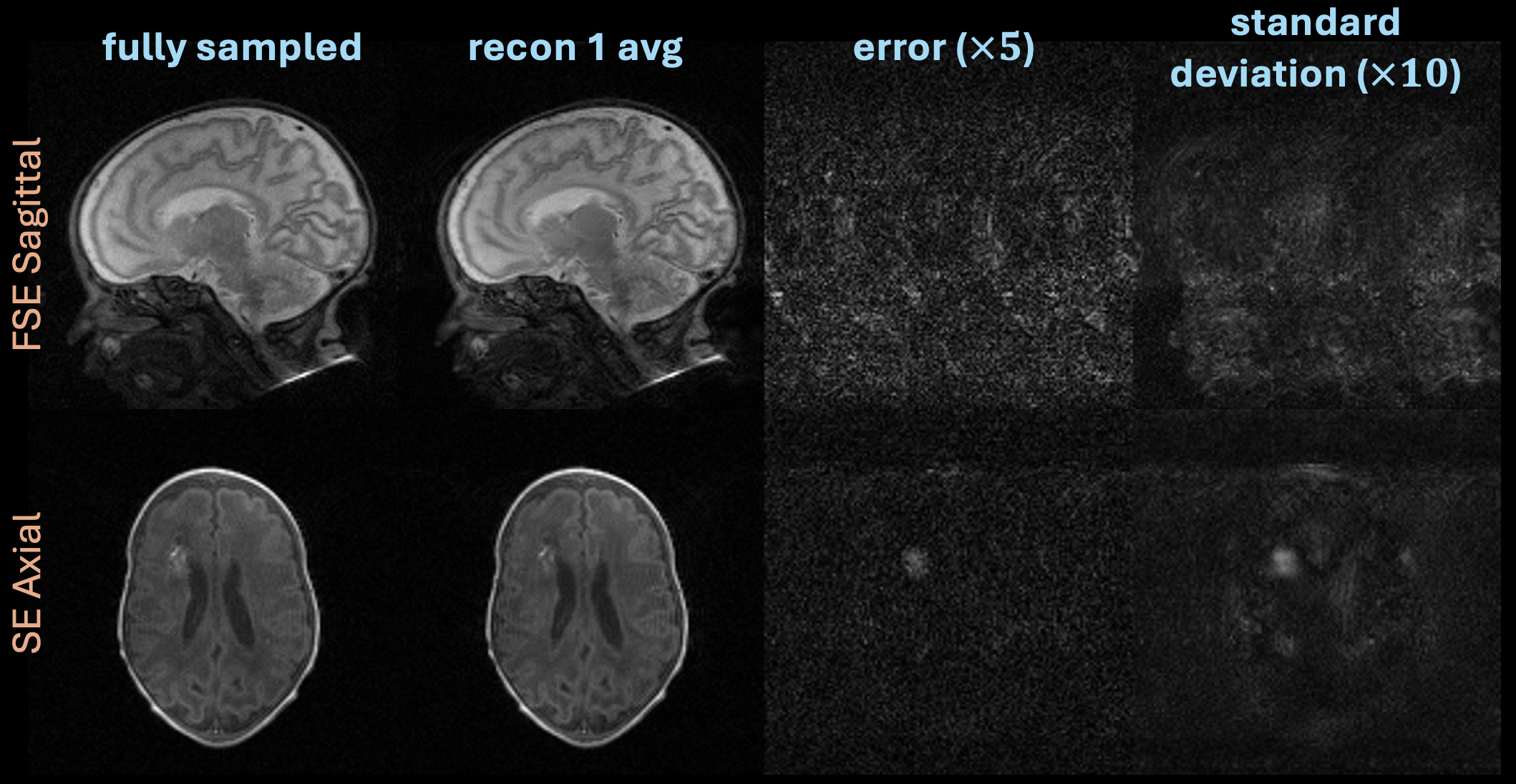}}
    \caption{Example FSE Sagittal and SE Axial fully-sampled images and associated reconstruction from one posterior sample. The corresponding error and standard deviation across multiple posterior samples show that reconstruction with posterior sampling using generative models enables potential estimates of reconstruction uncertainty by computing statistics over multiple posterior samples.}
\end{figure}
\renewcommand{\thefigure}{\arabic{figure}}  

\end{document}